\documentclass[preprint,aps,amsmath,nofootinbib,tightenlines,showpacs]{revtex4}
\usepackage{amsmath}
\usepackage{amsfonts}
\usepackage{amssymb}
\usepackage{latexsym}
\usepackage{graphicx}
\usepackage{bm}
\usepackage{epsfig}
\usepackage[english]{babel}

\def\sign{\mbox{sign}}

\begin{document}


\title{Note on Non-Gaussianities in Two-field Inflation}
\author{Tower Wang\footnote{Electronic address: twang@phy.ecnu.edu.cn}}
\affiliation{Department of Physics, East China Normal University,\\
Shanghai 200241, China\\ \vspace{0.2cm}}
\date{\today\\ \vspace{1cm}}
\begin{abstract}
Two-field slow-roll inflation is the most conservative modification
of a single-field model. The main motivations to study it are its
entropic mode and non-Gaussianity. Several years ago, for a
two-field model with additive separable potentials, Vernizzi and
Wands invented an analytic method to estimate its non-Gaussianities.
Later on, Choi \emph{et al}. applied this method to the model with
multiplicative separable potentials. In this note, we design a
larger class of models whose non-Gaussianity can be estimated by the
same method. Under some simplistic assumptions, roughly these models
are unlikely able to generate a large non-Gaussianity. We look over
some specific models of this class by scanning the full parameter
space, but still no large non-Gaussianity appears in the slow-roll
region. These models and scanning techniques would be useful for
future model hunt if observational evidence shows up for two-field
inflation.
\end{abstract}


\maketitle



\section{Introduction}\label{sect-intro}
Cosmic inflation \cite{Guth:1980zm} is a great idea to solve some
cosmological problems and to predict the fine fluctuations of cosmic
microwave background (CMB). Hitherto the surviving and most
economical model of inflation involves a single scalar field slowly
rolling down its effective potential
\cite{Linde:1981mu,Albrecht:1982wi}, with a canonical kinetic term
and minimally coupled to the Einstein gravity. We will call it the
simplest single-field inflation, although there is still freedom to
design its exact potential. The single-field inflation passed the
latest observational test \cite{Komatsu:2010fb} successfully, even
with the simplest quadratic potential.

Nevertheless there are perpetual attempts to modify the simplest
single-field inflation. Some of them are motivated by incorporating
inflation model into certain theoretical frameworks, such as the
standard model of particle physics
\cite{Bezrukov:2007ep,DeSimone:2008ei} or string theory
\cite{Kachru:2003sx}. Some others put their stake on signatures that
cannot appear in the simplest single-field model, such as a large
deviation from the Gaussian distribution in the CMB temperature
fluctuations.\footnote{As a partial list, see
\cite{Chingangbam:2009xi,Huang:2009xa,Gao:2009gd,Cai:2009hw,Huang:2009vk,Gao:2009fx,Chen:2009bc,Matsuda:2009np,Gao:2009at,Gao:2009qy,Enqvist:2009eq,Chen:2009zp,Gong:2009dh}
and references therein for various models and recent development
along this direction.}

Among these modifications, the two-field slow-roll inflation is the
most conservative one, at least in my personal point of view. It
introduces another scalar field rather than a non-conventional
Lagrangian such as non-canonical kinetic terms or modifications of
gravity. It also retains the slow-roll condition, which makes the
model simple and consistent with the observed CMB power spectrum. If
both conventional Lagrangian and non-conventional Lagrangian are
adaptable to the observational data, then the model with
conventional Lagrangian would be more acceptable, unless there are
better and solid theoretical motivations for non-conventional
Lagrangian.

On the observational side, two new features arise in two-field
model. First, the model is able to leave a residual entropic
perturbation between the fluctuations of dark matter and CMB
\cite{GarciaBellido:1995qq,Byrnes:2006fr}. Second, in a simple model
with quadratic potential, numerical computations
\cite{Rigopoulos:2005us,Vernizzi:2006ve} found that the
non-Gaussianity can be temporarily large at the turn of inflation
trajectory in field space. Longer-lived large non-Gaussianities were
discovered recently by
\cite{Byrnes:2008wi,Byrnes:2008zy,Byrnes:2009qy} in many other
two-field models.\footnote{The readers may refer to
\cite{Byrnes:2010em} for a review on this topic, and to
\cite{Bernardeau:2002jy,Bernardeau:2002jf} for pioneer works that
computed analytically the non-Gaussianity expected in multi-field
inflation. By studying the loop corrections,
\cite{Cogollo:2008bi,Rodriguez:2008hy} obtained an observable level of
non-Gaussianities, even when the two-field model is of the slow-roll
variety with canonical kinetic terms and in the framework of
Einstein gravity.}

Compared with the simplest one-field inflation, the field space
becomes two-dimensional in a two-field model. When the inflation
trajectory is curved in field space, the entropic perturbation will
be coupled to the adiabatic perturbation. So there are more
uncertainties in calculation of cosmological observables, such as
power spectra of CMB and their indices. It would be more complicated
to honestly compute the bispectra and non-linear parameters, which
reflect the non-Gaussianity of the primordial fluctuations.

Fortunately, based on the extended $\delta N$-formalism
\cite{Lyth:2005fi}, Vernizzi and Wands \cite{Vernizzi:2006ve}
invented an analytic method to estimate such non-Gaussianities. They
demonstrated the power of this method in a two-field model with
additive separable potentials. This method was later applied by Choi
\emph{et al}. \cite{Choi:2007su} to a model with multiplicative
separable potentials.

Encouraged by the method of Vernizzi and Wands, we tried to improve
it for the two-field slow-roll model with generic potentials but
failed. Finally, we only designed a larger class of models whose
non-Gaussianity can be estimated by this method. It is a class of
models whose potential take the form $W(w)$ with
$w=U(\varphi)+V(\chi)$ or $w=U(\varphi)V(\chi)$. Here $W(w)$,
$U(\varphi)$ and $V(\chi)$ are arbitrary functions of the indicated
variables as long as the slow-roll condition is satisfied. Scalar
fields $\varphi$ and $\chi$ are inflatons.

The outline of this paper is as follows. In our convention of
notations, we will prepare some well-known but necessary knowledge
in section \ref{sect-preparation} concisely. In section
\ref{sect-hunt}, we will present the exact form of our models, whose
non-linear parameters will be worked out in sections
\ref{sect-modelI} and \ref{sect-modelII}. Some specific examples are
investigated in section \ref{sect-examples}. We summarize the main
results of this paper in the final section.

This is a note concerning references
\cite{Vernizzi:2006ve,Choi:2007su}. Some of our techniques stem from
these references or slightly generalize theirs. Sometimes we employ
the techniques with few explanation if the mathematical
development is smooth. To better understand them, the readers are
strongly recommended to review the relevant parts of
\cite{Vernizzi:2006ve,Choi:2007su}.

\section{Non-Gaussianities in Two-field Inflation}\label{sect-preparation}
We are interested in inflation models described by the following
action \cite{DiMarco:2005nq,Choi:2007su}
\begin{equation}\label{action}
S=\int d^4x\sqrt{-g}\left[\frac{1}{2}M_p^2R-\frac{1}{2}g^{\mu\nu}\partial_{\mu}\varphi\partial_{\nu}\varphi-\frac{1}{2}e^{2b(\varphi)}g^{\mu\nu}\partial_{\mu}\chi\partial_{\nu}\chi-W(\varphi,\chi)\right].
\end{equation}
Because of the appearance of $b(\varphi)$, the field $\chi$ has a
non-standard kinetic term. Following the notation of slow-roll
parameters defined in \cite{GarciaBellido:1995qq,DiMarco:2005nq}
\begin{eqnarray}\label{slroll}
\nonumber &&\epsilon_{\varphi}=\frac{M_p^2}{2}\left(\frac{W_{,\varphi}}{W}\right)^2,~~~~\epsilon_{\chi}=\frac{M_p^2}{2}\left(\frac{W_{,\chi}}{W}\right)^2e^{-2b},~~~~\epsilon_b=8M_p^2b_{,\varphi}^2,\\
&&\eta_{\varphi\varphi}=\frac{M_p^2W_{,\varphi\varphi}}{W},~~~~\eta_{\varphi\chi}=\frac{M_p^2W_{,\varphi\chi}}{W}e^{-b},~~~~\eta_{\chi\chi}=\frac{M_p^2W_{,\chi\chi}}{W}e^{-2b},
\end{eqnarray}
the slow-roll condition can be expressed as $\epsilon_i\ll1$,
$\epsilon_b\ll1$, $|\eta_{ij}|\ll1$ with $i,j=\varphi,\chi$.

As an aside, we mention that model \eqref{action} is equivalent to
the $f(\chi,R)$ generalized gravity \cite{Hwang:2005hb,Ji:2009yw}
when $b=-\varphi/(\sqrt{6}M_p)$. But then we find $\epsilon_b=4/3$,
which violates the the slow-roll condition. This is a pitfall in
treating generalized gravity as a two-field model. This pitfall can
be circumvented by the scheme in \cite{Ji:2009yw}.

Under the slow-roll condition, the background equations of motion
are very simple
\begin{equation}\label{eom}
3H\dot{\varphi}+W_{,\varphi}=0,~~~~3He^{2b}\dot{\chi}+W_{,\chi}=0,~~~~3M_p^2H^2=W.
\end{equation}
Using them one may directly demonstrate
\begin{equation}
\epsilon=-\frac{\dot{H}}{H^2}=\epsilon_{\varphi}+\epsilon_{\chi}.
\end{equation}

Observationally, the most promising probe of primordial
non-Gaussianities comes from the bispectrum of CMB fluctuations,
which is characterized by the non-linear parameter
$f_{\mathrm{NL}}(\mathbf{k}_1,\mathbf{k}_2,\mathbf{k}_3)$. If
$|f_{\mathrm{NL}}|\gtrsim10$, it would be detectable by ongoing or
planned satellite experiments \cite{planck,cmbpol}.

It has been shown in \cite{Seery:2005gb,Vernizzi:2006ve,Choi:2007su}
that the non-linear parameter in two-field inflation models can be
separated into a momentum dependent term and a momentum independent
term
\begin{equation}\label{fNL}
-\frac{6}{5}f_{\mathrm{NL}}(\mathbf{k}_1,\mathbf{k}_2,\mathbf{k}_3)=-\frac{6}{5}f_{\mathrm{NL}}^{(3)}(\mathbf{k}_1,\mathbf{k}_2,\mathbf{k}_3)-\frac{6}{5}f_{\mathrm{NL}}^{(4)}
\end{equation}
It is also proved in \cite{Vernizzi:2006ve,Choi:2007su} that the
first term is always suppressed by the tensor-to-scalar ratio,
leading to $|f_{\mathrm{NL}}^{(3)}|\ll1$. Hence this term is
negligible in observation. For action \eqref{action}, the second
term
\begin{equation}\label{fNL4}
-\frac{6}{5}f_{\mathrm{NL}}^{(4)}=\frac{N_{,\varphi_*}^2N_{,\varphi_*\varphi_*}+2e^{-2b_*}N_{,\varphi_*}N_{,\chi_*}N_{,\varphi_*\chi_*}+e^{-4b_*}N_{,\chi_*}^2N_{,\chi_*\chi_*}}{\left(N_{,\varphi_*}^2+e^{-2b_*}N_{,\chi_*}^2\right)^2}
\end{equation}
may be large and deserves a closer look. Here $N=\int_*^cHdt$ is the
$e$-folding number from the initial flat hypersurface $t=t_*$ to the
final comoving hypersurface $t=t_c$. To evaluate \eqref{fNL4}, we
will work out the derivatives of $N$ with respect to $\varphi_*$ and
$\chi_*$ in the next section, focusing on a class of analytically
solvable models.

\section{Hunting for Analytically Solvable Models}\label{sect-hunt}
Making use of equations \eqref{eom}, the $e$-folding number can be
cast as
\begin{eqnarray}\label{N}
\nonumber N&=&\int_*^cHdt\\
\nonumber &=&-\int_*^c\frac{(W-Q)\dot{\varphi}}{M_p^2W_{,\varphi}}dt-\int_*^c\frac{Q(\varphi,\chi)\dot{\chi}}{M_p^2W_{,\chi}}e^{2b(\varphi)}dt\\
&=&-\int_*^c\frac{W-Q}{M_p^2W_{,\varphi}}d\varphi-\int_*^c\frac{Qe^{2b}}{M_p^2W_{,\chi}}d\chi.
\end{eqnarray}
Hence $Q(\varphi,\chi)$ is an arbitrary function of $\varphi$ and
$\chi$ in principle, because
$\dot{\varphi}/W_{,\varphi}=e^{2b}\dot{\chi}/W_{,\chi}=-1/(3H)$
along any classical trajectory under the slow-roll condition.
However, for a given $W$, we have to choose a suitable form of $Q$
so that the integrations defined by $\mathcal{Q}$ in \eqref{ZQ} can
be performed. Later on we will fix $Q$ to meet the ansatz
\eqref{ansatzQ} for simplicity. But for the moment let us leave it
as an arbitrary function of $\varphi$ and $\chi$. It is
straightforward to obtain the first order partial derivatives
\begin{eqnarray}\label{dN}
\nonumber \frac{\partial N}{\partial\varphi_*}&=&\frac{W^*-Q^*}{M_p^2W^*_{,\varphi}}-\frac{W^c-Q^c}{M_p^2W^c_{,\varphi}}\frac{\partial\varphi_c}{\partial\varphi_*}-\int_*^c\left(\frac{W-Q}{M_p^2W_{,\varphi}}\right)_{,\chi}\frac{\partial\chi}{\partial\varphi_*}d\varphi\\
\nonumber &&-\frac{Q^ce^{2b_c}}{M_p^2W^c_{,\chi}}\frac{\partial\chi_c}{\partial\varphi_*}-\int_*^c\left(\frac{Qe^{2b}}{M_p^2W_{,\chi}}\right)_{,\varphi}\frac{\partial\varphi}{\partial\varphi_*}d\chi,\\
\nonumber \frac{\partial N}{\partial\chi_*}&=&-\frac{W^c-Q^c}{M_p^2W^c_{,\varphi}}\frac{\partial\varphi_c}{\partial\chi_*}-\int_*^c\left(\frac{W-Q}{M_p^2W_{,\varphi}}\right)_{,\chi}\frac{\partial\chi}{\partial\chi_*}d\varphi\\
&&+\frac{Q^*e^{2b_*}}{M_p^2W^*_{,\chi}}-\frac{Q^ce^{2b_c}}{M_p^2W^c_{,\chi}}\frac{\partial\chi_c}{\partial\chi_*}-\int_*^c\left(\frac{Qe^{2b}}{M_p^2W_{,\chi}}\right)_{,\varphi}\frac{\partial\varphi}{\partial\chi_*}d\chi.
\end{eqnarray}

Akin to \cite{Vernizzi:2006ve,Choi:2007su}, we define an integral of
motion $C$ along the trajectory of inflation
\begin{equation}\label{C}
C=\int\frac{Wf(\varphi,\chi)}{M_p^2W_{,\varphi}}d\varphi-\int\frac{We^{2b}f}{M_p^2W_{,\chi}}d\chi,
\end{equation}
Here the explicit form of $f(\varphi,\chi)$ is determined by scalar
potential $W(\varphi,\chi)$. We will give the expression of $f$ for
some types of potential in this section. If we fix the limits of
integration to run from $t_*$ to $t_c$, then due to the background
equations \eqref{eom},
\begin{equation}\label{Cstarc}
\left.C\right|_*^c=\int_*^c\frac{Wf}{3M_p^2H}\left(\frac{\dot{\varphi}}{W_{,\varphi}}-\frac{e^{2b}\dot{\chi}}{W_{,\chi}}\right)dt=0
\end{equation}
along classical trajectories under the slow-roll approximation. So
the constant $C$ parameterizes the motion off classical
trajectories. In order to know
$\partial\varphi_c/\partial\varphi_*$,
$\partial\chi_c/\partial\varphi_*$,
$\partial\varphi_c/\partial\chi_*$, $\partial\chi_c/\partial\chi_*$
in \eqref{dN}, we should calculate the first order derivatives of
$C$ on the initial flat hypersurface $t=t_*$,
\begin{eqnarray}\label{dC}
\nonumber \frac{\partial C}{\partial\varphi_*}&=&\frac{W^*f^*}{M_p^2W^*_{,\varphi}}+\int^*\left(\frac{Wf}{M_p^2W_{,\varphi}}\right)_{,\chi}\frac{\partial\chi}{\partial\varphi_*}d\varphi-\int^*\left(\frac{We^{2b}f}{M_p^2W_{,\chi}}\right)_{,\varphi}\frac{\partial\varphi}{\partial\varphi_*}d\chi,\\
\frac{\partial C}{\partial\chi_*}&=&\int^*\left(\frac{Wf}{M_p^2W_{,\varphi}}\right)_{,\chi}\frac{\partial\chi}{\partial\chi_*}d\varphi-\frac{W^*e^{2b_*}f^*}{M_p^2W^*_{,\chi}}-\int^*\left(\frac{We^{2b}f}{M_p^2W_{,\chi}}\right)_{,\varphi}\frac{\partial\varphi}{\partial\chi_*}d\chi.
\end{eqnarray}
Differentiating \eqref{C} with respect to $C$, it gives
\begin{equation}\label{dCdC}
1=\frac{Wf}{M_p^2W_{,\varphi}}\frac{d\varphi}{dC}+\int\left(\frac{Wf}{M_p^2W_{,\varphi}}\right)_{,\chi}\frac{d\chi}{dC}d\varphi-\frac{We^{2b}f}{M_p^2W_{,\chi}}\frac{d\chi}{dC}-\int\left(\frac{We^{2b}f}{M_p^2W_{,\chi}}\right)_{,\varphi}\frac{d\varphi}{dC}d\chi.
\end{equation}

On large scales, the comoving hypersurface $t=t_c$ coincides with
the uniform density hypersurface. This implies under the slow-roll
condition
\begin{equation}\label{condens}
W(\varphi_c,\chi_c)=\mathrm{const.},
\end{equation}
whose differentiation with respect to $C$ is
\begin{equation}\label{dcondensdC}
W^c_{,\varphi}\frac{d\varphi_c}{dC}+W^c_{,\chi}\frac{d\chi_c}{dC}=0.
\end{equation}
Combined with \eqref{dCdC} on the final comoving surface $t=t_c$, it
could give the solution for $d\varphi_c/dC$ and $d\chi_c/dC$. This
is in general difficult analytically. To overcome the difficulty, we
introduce an ansatz:
\begin{equation}\label{ansatzf}
\left(\frac{Wf}{M_p^2W_{,\varphi}}\right)_{,\chi}=\left(\frac{We^{2b}f}{M_p^2W_{,\chi}}\right)_{,\varphi}=0.
\end{equation}

Although we are free to design the function $f(\varphi,\chi)$, the
above condition is not always satisfiable. We have hunted for
analytical models meeting this condition, and found it is achievable
if $W=W(w)$ with $w=U(\varphi)+V(\chi)$ or $w=U(\varphi)V(\chi)$.
Here $W(w)$, $U(\varphi)$ and $V(\chi)$ are arbitrary functions of
the indicated variables as long as the slow-roll condition is
satisfied. In this paper, we will pay attention to this
situation. But it is never excluded that there might be other
situations in which $d\varphi_c/dC$ and $d\chi_c/dC$ are solvable
from \eqref{dCdC} and \eqref{dcondensdC}, even if ansatz
\eqref{ansatzf} is violated.

Ansatz \eqref{ansatzf} simplifies our discussion significantly. Once
it holds, equations \eqref{dCdC} and \eqref{dcondensdC} lead to
\begin{eqnarray}\label{dC1}
\nonumber \frac{d\varphi_c}{dC}&=&\frac{2W^c}{W^c_{,\varphi}f^c}\frac{\epsilon^c_{\varphi}\epsilon^c_{\chi}}{\epsilon^c},\\
\frac{d\chi_c}{dC}&=&-\frac{2W^c}{W^c_{,\chi}f^c}\frac{\epsilon^c_{\varphi}\epsilon^c_{\chi}}{\epsilon^c},
\end{eqnarray}
while \eqref{dC} is reduced as
\begin{eqnarray}\label{dC2}
\nonumber \frac{\partial C}{\partial\varphi_*}&=&\frac{W^*f^*}{M_p^2W^*_{,\varphi}},\\
\frac{\partial C}{\partial\chi_*}&=&-\frac{W^*e^{2b_*}f^*}{M_p^2W^*_{,\chi}}.
\end{eqnarray}

As a result, the partial derivatives of $N$ take the form
\begin{eqnarray}
\nonumber N_{,\varphi_*}&=&\frac{W^*-Q^*}{M_p^2W^*_{,\varphi}}+(Z^c-\mathcal{Q})\frac{\partial C}{\partial\varphi_*},\\
\nonumber N_{,\chi_*}&=&\frac{Q^*e^{2b_*}}{M_p^2W^*_{,\chi}}+(Z^c-\mathcal{Q})\frac{\partial C}{\partial\chi_*},\\
\nonumber N_{,\varphi_*\varphi_*}&=&\left(\frac{W^*-Q^*}{M_p^2W^*_{,\varphi}}\right)_{,\varphi_*}+\left(\frac{\partial Z^c}{\partial\varphi_*}-\frac{\partial\mathcal{Q}}{\partial\varphi_*}\right)\frac{\partial C}{\partial\varphi_*}+(Z^c-\mathcal{Q})\frac{\partial^2C}{\partial\varphi_*^2},\\
\nonumber N_{,\varphi_*\chi_*}&=&\left(\frac{W^*-Q^*}{M_p^2W^*_{,\varphi}}\right)_{,\chi_*}+\left(\frac{\partial Z^c}{\partial\chi_*}-\frac{\partial\mathcal{Q}}{\partial\chi_*}\right)\frac{\partial C}{\partial\varphi_*}+(Z^c-\mathcal{Q})\frac{\partial^2C}{\partial\varphi_*\partial\chi_*},\\
N_{,\chi_*\chi_*}&=&\left(\frac{Q^*e^{2b_*}}{M_p^2W^*_{,\chi}}\right)_{,\chi_*}+\left(\frac{\partial Z^c}{\partial\chi_*}-\frac{\partial\mathcal{Q}}{\partial\chi_*}\right)\frac{\partial C}{\partial\chi_*}+(Z^c-\mathcal{Q})\frac{\partial^2C}{\partial\chi_*^2}.
\end{eqnarray}
In these equations, we have adopted the notations
\begin{eqnarray}\label{ZQ}
\nonumber Z^c&=&\frac{Q^c\epsilon^c_{\varphi}-(W^c-Q^c)\epsilon^c_{\chi}}{W^cf^c\epsilon^c},\\
\nonumber \mathcal{Q}&=&\int_*^c\left(\frac{W-Q}{M_p^2W_{,\varphi}}\right)_{,\chi}\frac{d\chi}{dC}d\varphi+\int_*^c\left(\frac{Qe^{2b}}{M_p^2W_{,\chi}}\right)_{,\varphi}\frac{d\varphi}{dC}d\chi,\\
\nonumber \frac{\partial Z^c}{\partial\varphi_*}&=&Z^c_{,\varphi}\frac{d\varphi_c}{dC}\frac{\partial C}{\partial\varphi_*}+Z^c_{,\chi}\frac{d\chi_c}{dC}\frac{\partial C}{\partial\varphi_*},\\
\nonumber \frac{\partial\mathcal{Q}}{\partial\varphi_*}&=&\left(\frac{W^c-Q^c}{M_p^2W^c_{,\varphi}}\right)_{,\chi_c}\frac{d\chi_c}{dC}\frac{\partial\varphi_c}{\partial\varphi_*}-\left(\frac{W^*-Q^*}{M_p^2W^*_{,\varphi}}\right)_{,\chi_*}\frac{d\chi_*}{dC}\\
\nonumber &&+\frac{\partial C}{\partial\varphi_*}\int_*^c\left(\frac{W-Q}{M_p^2W_{,\varphi}}\right)_{,\chi\chi}\left(\frac{d\chi}{dC}\right)^2d\varphi+\frac{\partial C}{\partial\varphi_*}\int_*^c\left(\frac{W-Q}{M_p^2W_{,\varphi}}\right)_{,\chi}\frac{d^2\chi}{dC^2}d\varphi\\
\nonumber &&+\left(\frac{Q^ce^{2b_c}}{M_p^2W^c_{,\chi}}\right)_{,\varphi_c}\frac{d\varphi_c}{dC}\frac{\partial\chi_c}{\partial\varphi_*}+\frac{\partial C}{\partial\varphi_*}\int_*^c\left(\frac{Qe^{2b}}{M_p^2W_{,\chi}}\right)_{,\varphi\varphi}\left(\frac{d\varphi}{dC}\right)^2d\chi\\
&&+\frac{\partial C}{\partial\varphi_*}\int_*^c\left(\frac{Qe^{2b}}{M_p^2W_{,\chi}}\right)_{,\varphi}\frac{d^2\varphi}{dC^2}d\chi.
\end{eqnarray}

In the above, the expression of $\mathcal{Q}$ and its derivatives
involve nuisance integrals. To further simplify our study, we
utilize one more ansatz
\begin{equation}\label{ansatzQ}
\left(\frac{W-Q}{M_p^2W_{,\varphi}}\right)_{,\chi}=\left(\frac{Qe^{2b}}{M_p^2W_{,\chi}}\right)_{,\varphi}=0.
\end{equation}
In favor of this ansatz, we have $\mathcal{Q}=0$ and so do its
derivatives.

As was mentioned, ansatz \eqref{ansatzf} can be satisfied by special
forms of potential $W(\varphi,\chi)$. Now ansatz \eqref{ansatzQ}
further constrains the form of $W(\varphi,\chi)$ and $b(\varphi)$.
Let us discuss it in details case by case.

\subsection{Case I: $W=W(w)$, $w=U(\varphi)+V(\chi)$}\label{subsect-caseI}
For this class of models, according to \eqref{ansatzf}, we set
\begin{equation}\label{fI}
f=\frac{W_{,w}}{We^{2b}},
\end{equation}
while condition \eqref{ansatzQ} is met by
\begin{equation}\label{modelIa}
b=0,~~~~W=\lambda w^{\alpha},~~~~Q=\lambda Vw^{\alpha-1}
\end{equation}
or
\begin{equation}\label{modelIb}
b=-\frac{1}{2}\nu U,~~~~\frac{d\ln W}{dw}=\frac{1}{p+qe^{\nu w}},~~~~Q=qe^{\nu w}W_{,w}.
\end{equation}
Hereafter, as free parameters in our models, $\lambda$, $\alpha$,
$\beta$, $\nu$, $p$ and $q$ are arbitrary real constants. The
normalization of $e^{2b}$ is fixed for simplicity. This is always
realizable by rescaling the field $\chi$.

Taking $\alpha=1$, model \eqref{modelIa} recovers the well-studied
sum potential
\cite{Polarski:1992dq,Langlois:1999dw,Vernizzi:2006ve}, to which we
will return in subsection \ref{subsect-sumPot}. In subsection
\ref{subsect-nsPotII}, we will study a specific example of
non-separable potential that corresponds to $\alpha=2$ in
\eqref{modelIa}.

As will be discussed in subsection \ref{subsect-IeqII}, there is an
equivalence relation between case I in this subsection and case II
in the next subsection. Models in class I can be transformed to
those in class II, and \emph{vice versa}. We will translate model
\eqref{modelIb} to a nicer form \eqref{modelIIb} and explore it.

\subsection{Case II: $W=W(w)$, $w=U(\varphi)V(\chi)$}\label{subsect-caseII}
For this class of models, we take
\begin{equation}\label{fII}
f=\frac{wW_{,w}}{We^{2b}},
\end{equation}
then condition \eqref{ansatzf} is satisfied. Condition
\eqref{ansatzQ} can be met by
\begin{equation}\label{modelIIa}
b=0,~~~~W=\lambda (\ln w)^{\alpha},~~~~Q=\lambda(\ln w)^{\alpha-1}\ln V
\end{equation}
or
\begin{equation}\label{modelIIb}
e^{2b}=U^{-\nu},~~~~\frac{d\ln W}{dw}=\frac{1}{pw+qw^{\nu+1}},~~~~Q=qw^{\nu+1}W_{,w}.
\end{equation}
We observed that \eqref{fII}, \eqref{modelIIa} and \eqref{modelIIb}
can be obtained from \eqref{fI}, \eqref{modelIa} and \eqref{modelIb}
perfectly by the following replacement:
\begin{equation}\label{ItoII}
U\rightarrow\ln U,~~~~V\rightarrow\ln V,~~~~w\rightarrow\ln w.
\end{equation}
In fact, there is a general equivalence relation between case I and
case II, on which will be elaborated in subsection
\ref{subsect-IeqII}.

Equation \eqref{modelIIb} dictates $W$ implicitly as a differential
equation. To obtain the explicit form of $W$, one should solve the
equation. This could be done analytically in some corners of the
parameter space. For instance, setting $\nu=0$, equation
\eqref{modelIIb} gives
\begin{equation}\label{modelIIb1}
b=0,~~~~W=\lambda w^{\alpha},~~~~Q=\beta w^{\alpha}.
\end{equation}
However, if $q=0$, it leads to a larger class of model
\begin{equation}\label{modelIIb2}
W=\lambda w^{\alpha},~~~~Q=0,
\end{equation}
leaving $b$ as an arbitrary function of $\varphi$. Model
\eqref{modelIIb1} or \eqref{modelIIb2} is separable and can be seen
as the well-studied product potential
\cite{GarciaBellido:1995qq,Choi:2007su}. More discussion on models
with product potential will be given in subsection
\ref{subsect-prodPot}. In the case that $p=0$ and $\nu\neq0$, we
find another model
\begin{equation}\label{modelIIb3}
e^{2b}=U^{\alpha},~~~~W=Q=\lambda\exp\left(\beta w^{\alpha}\right).
\end{equation}
In subsection \ref{subsect-nsPotII}, we will study an example of
non-separable potential which corresponds to $\alpha=1$ in
\eqref{modelIIb3}. Since $\nu$ is an arbitrary real constant,
equation \eqref{modelIIb} can generate many other forms of potential
$W$. For example, when $p\neq0$ and $\nu=-1$, we get a model
\begin{equation}\label{modelIIb4}
e^{2b}=U,~~~~W=\lambda(w+\beta)^{\alpha},~~~~Q=\lambda\beta(w+\beta)^{\alpha-1}.
\end{equation}

\subsection{Equivalence between Case I and Case II}\label{subsect-IeqII}
We have classified our models into two categories, corresponding to
subsections \ref{subsect-caseI} and \ref{subsect-caseII}. In case I,
the potential $W(w)$ is a function of sum $w=U(\varphi)+V(\chi)$. In
case II, the potential $W(w)$ is a function of product
$w=U(\varphi)V(\chi)$. After the non-dimensionalization, case I can
be translated to case II by the transformation
\begin{equation}\label{IItoI}
U\rightarrow e^U,~~~~V\rightarrow e^V,~~~~w\rightarrow e^w.
\end{equation}
The last relation in \eqref{IItoI} is a corollary of the former ones
because $UV\rightarrow e^{U+V}$. On the other hand, via
transformation \eqref{ItoII}, an arbitrary potential of case I can
be transformed to that of case II. So the two ``cases''are just two
different formalisms for studying the same models. They are
equivalent to each other. We are free to study a model in either
formalism contingent on the convenience.

For instance, using the formulae in this section, a model with
potential $W=\lambda e^{-\beta\varphi^2\chi^2}$ and prefactor
$e^{2b}=\alpha\varphi^2$ can be studied in two different formalisms:
\begin{itemize}
\item Formalism I: $W=\lambda\exp(-e^w)$, $b=U/2$ with $w=U+V$, $U=\ln(\alpha\varphi^2)$,
$V=\ln(\beta\chi^2/\alpha)$.
\item Formalism II: $W=\lambda e^{-w}$, $e^{2b}=U$ with $w=UV$, $U=\alpha\varphi^2$,
$V=\beta\chi^2/\alpha$.
\end{itemize}
But apparently, for this model the calculation will be easier in
formalism II. Because the dependence of $W$ and $b$ on $\varphi$ and
$\chi$ is unaltered, the quantization of perturbations is not
affected by the choice of formalism. For the same reason, the exact
dependence of $f_{NL}$ on $\varphi$ and $\chi$ is the same in both
formalisms.

\section{Model I: $W=\lambda w^{\alpha}$, $w=U(\varphi)+V(\chi)$, $b=0$}\label{sect-modelI}
This model is given by \eqref{modelIa}, which is equivalent to model
\eqref{modelIIa}. Corresponding to this model, the number of
$e$-foldings and the integral constant along the inflation
trajectory are
\begin{eqnarray}
\nonumber N&=&-\frac{1}{\alpha M_p^2}\left(\int_*^c\frac{U}{U_{,\varphi}}d\varphi+\int_*^c\frac{V}{V_{,\chi}}d\chi\right),\\
C&=&\frac{1}{M_p^2}\left(\int\frac{d\varphi}{U_{,\varphi}}-\int\frac{d\chi}{V_{,\chi}}\right).
\end{eqnarray}

We have defined the slow-roll parameters in \eqref{slroll}. In the
present case, they are of the form
\begin{eqnarray}\label{slrollI}
\nonumber &&\epsilon_{\varphi}=\frac{\alpha^2M_p^2}{2}\frac{U_{,\varphi}^2}{w^2},~~~~\epsilon_{\chi}=\frac{\alpha^2M_p^2}{2}\frac{V_{,\chi}^2}{w^2},\\
\nonumber &&\epsilon_b=0,~~~~\epsilon=\frac{\alpha^2M_p^2}{2}\frac{U_{,\varphi}^2+V_{,\chi}^2}{w^2},\\
\nonumber &&\eta_{\varphi\varphi}=\frac{\alpha(\alpha-1)}{w^2}M_p^2U_{,\varphi}^2+\frac{\alpha}{w}M_p^2U_{,\varphi\varphi},\\
\nonumber &&\eta_{\varphi\chi}=\frac{\alpha(\alpha-1)}{w^2}M_p^2U_{,\varphi}V_{,\chi},\\
&&\eta_{\chi\chi}=\frac{\alpha(\alpha-1)}{w^2}M_p^2V_{,\chi}^2+\frac{\alpha}{w}M_p^2V_{,\chi\chi}.
\end{eqnarray}

Now equations \eqref{dC1} and \eqref{dC2} become
\begin{eqnarray}
\nonumber \frac{d\varphi_c}{dC}=\left.\frac{M_p^2U_{,\varphi}V_{,\chi}^2}{U_{,\varphi}^2+V_{,\chi}^2}\right|_c,&&\frac{d\chi_c}{dC}=-\left.\frac{M_p^2U_{,\varphi}^2V_{,\chi}}{U_{,\varphi}^2+V_{,\chi}^2}\right|_c,\\
\frac{\partial C}{\partial\varphi_*}=\frac{1}{M_p^2U^*_{,\varphi}},&&\frac{\partial C}{\partial\chi_*}=-\frac{1}{M_p^2V^*_{,\chi}},
\end{eqnarray}
while the function $Z$ defined by \eqref{ZQ} takes the form
\begin{eqnarray}
&&Z=\frac{U_{,\varphi}^2V-UV_{,\chi}^2}{\alpha(U_{,\varphi}^2+V_{,\chi}^2)}.
\end{eqnarray}
Then we get the partial derivatives of $Z^c$ with respect to
$\varphi_*$ and $\chi_*$,
\begin{eqnarray}
&&\frac{\alpha M_pU^*_{,\varphi}}{\sqrt{2}w^*}\frac{\partial Z^c}{\partial\varphi_*}=-\frac{\alpha M_pV^*_{,\chi}}{\sqrt{2}w^*}\frac{\partial Z^c}{\partial\chi_*}=\frac{\sqrt{2}w^*\mathcal{A}}{M_p}\
\end{eqnarray}
in terms of
\begin{eqnarray}
&&\mathcal{A}=-\frac{w^{c2}}{w^{*2}}\frac{\epsilon^c_{\varphi}\epsilon^c_{\chi}}{\alpha^2\epsilon^c}\left[1+\frac{4(\alpha-1)\epsilon^c_{\varphi}\epsilon^c_{\chi}}{\epsilon^{c2}}-\frac{\alpha(\epsilon^c_{\chi}\eta^c_{\varphi\varphi}+\epsilon^c_{\varphi}\eta^c_{\chi\chi})}{\epsilon^{c2}}\right].
\end{eqnarray}

With the above result at hand, it is straightforward to calculate
\begin{eqnarray}
\nonumber &&N_{,\varphi_*}=\frac{w^*u}{\alpha M_p^2U^*_{,\varphi}},~~~~N_{,\chi_*}=\frac{w^*v}{\alpha M_p^2V^*_{,\chi}},\\
\nonumber &&N_{,\varphi_*\varphi_*}=\frac{1}{\alpha M_p^2}\left[\left(1-\frac{\eta^*_{\varphi\varphi}}{2\epsilon^*_{\varphi}}\right)\alpha u+v+\frac{\alpha^2}{\epsilon^*_{\varphi}}\mathcal{A}\right],\\
\nonumber &&N_{,\varphi_*\chi_*}=-\frac{2w^{*2}\mathcal{A}}{\alpha M_p^4U^*_{,\varphi}V^*_{,\chi}},\\
&&N_{,\chi_*\chi_*}=\frac{1}{\alpha M_p^2}\left[\left(1-\frac{\eta^*_{\chi\chi}}{2\epsilon^*_{\chi}}\right)\alpha v+u+\frac{\alpha^2}{\epsilon^*_{\chi}}\mathcal{A}\right],
\end{eqnarray}
where for convenience we used notations
\begin{equation}
u=\frac{U^*+\alpha Z^c}{w^*},~~~~v=\frac{V^*-\alpha Z^c}{w^*}.
\end{equation}
For these notations, the relation $u+v=1$ holds. In the next
section, the definitions of $u$ and $v$ are different, but the same
relation also holds.

As a result, using formula \eqref{fNL4} we get the main part of
non-linear parameter in this model
\begin{eqnarray}\label{fNL4I}
\nonumber -\frac{6}{5}f_{\mathrm{NL}}^{(4)}&=&\frac{2}{\alpha}\left(\frac{u^2}{\epsilon^*_{\varphi}}+\frac{v^2}{\epsilon^*_{\chi}}\right)^{-2}\biggl\{\frac{u^2}{\epsilon^*_{\varphi}}\left[\left(1-\frac{\eta^*_{\varphi\varphi}}{2\epsilon^*_{\varphi}}\right)\alpha u+v\right]\\
&&+\frac{v^2}{\epsilon^*_{\chi}}\left[\left(1-\frac{\eta^*_{\chi\chi}}{2\epsilon^*_{\chi}}\right)\alpha v+u\right]+\left(\frac{u}{\epsilon^*_{\varphi}}+\frac{v}{\epsilon^*_{\chi}}\right)^2\alpha^2\mathcal{A}\biggr\}.
\end{eqnarray}

The non-linear parameter \eqref{fNL4I} depends on the exponent
$\alpha$ in a complicated manner. For the purpose of rough
estimation, we assume both $u$ and $v$ are of order unity. This
assumption is reasonable if $U^*$, $V^*$ and $w^*$ are of the same
order. It is also consistent with the relation $u+v=1$. Furthermore,
motivated by the slow-roll condition and the observational
constraint on spectral indices, we assume the slow-roll parameters
are of order $\mathcal{O}(10^{-2})$. In saying this we mean all of
the slow-roll parameters are of the same order, which is a strong
but still allowable assumption. After making these assumptions, we
can estimate the magnitude of \eqref{fNL4I} in three regions
according to the value of $\alpha$.

Firstly, in the limit $\alpha\ll1$, we have
$\alpha^2\mathcal{A}\sim\epsilon w^{c2}/w^{*2}\sim\epsilon$. So the
third term in curly brackets of \eqref{fNL4I} is of order
$\epsilon^{-1}w^{c2}/w^{*2}$, while the other two terms are of order
$\epsilon^{-1}$. Consequently, we can estimate
$f_{\mathrm{NL}}^{(4)}\sim\epsilon/\alpha$. It seems that a small
value of $\alpha$ could give rise to a large non-linear parameter.
Specifically, under our assumptions above, if
$\alpha\sim\mathcal{O}(10^{-3})$, then the non-linear parameter
$f_{\mathrm{NL}}\sim\mathcal{O}(10)$. However, this limit violates
our assumptions. On the one hand, we have assumed
$\epsilon_{\varphi}\sim\epsilon_{\chi}\sim\eta_{\varphi\chi}$. On
the other hand, equations \eqref{slrollI} tell us
$\epsilon_{\varphi}\epsilon_{\chi}/\eta_{\varphi\chi}^2\sim\alpha^2/(\alpha-1)^2$,
which apparently violates our assumption in the limit $\alpha\ll1$.
So we cannot use the oversimplified assumptions to estimate the
non-linear parameter in this limit.

Secondly, for $\alpha\gg1$, we would have
$\alpha^2\mathcal{A}\sim\alpha\epsilon
w^{c2}/w^{*2}\sim\alpha\epsilon$. Then the last term in braces of
\eqref{fNL4I} is of order $\alpha\epsilon^{-1}w^{c2}/w^{*2}$. The
other terms can be of order $\alpha\epsilon^{-1}$. After cancelation
with the prefactor, it leads to the estimation
$f_{\mathrm{NL}}^{(4)}\sim\epsilon$. That is to say, in this limit,
the non-linear parameter is independent of $\alpha$ in the leading
order and suppressed by the slow-roll parameters.

The third region is $\alpha\sim\mathcal{O}(1)$. In this region, the
non-linear parameter is still suppressed,
$f_{\mathrm{NL}}^{(4)}\sim\epsilon$.

Our conclusion is somewhat unexciting. This model could not generate
large non-Gaussianities under our simplistic assumptions. However,
one should be warned that our estimation above relies on two
assumptions: $u\sim v\sim\mathcal{O}(1)$ and $\epsilon\sim\eta\ll1$.
Although these assumptions are reasonable, they may be avoided in
very special circumstances. To further look for a large
non-Gaussianity with our formula \eqref{fNL4I}, one should give up
these assumptions and carefully scan the whole parameter space in a
consistent way. Generally that is an ambitious task if not
impossible. But  for a specific model of this type, we will perform
such a scanning in subsection \ref{subsect-nsPotII}.

\section{Model II: $d\ln W/dw=(pw+qw^{\nu+1})^{-1}$, $w=U(\varphi)V(\chi)$, $e^{2b}=U^{-\nu}$}\label{sect-modelII}
As we have discussed, model \eqref{modelIb} and model
\eqref{modelIIb} are equivalent. Thus it is enough to study them in
the relatively simpler form, namely in the form \eqref{modelIIb}.
For this model, we calculated the number of $e$-foldings and the
integral constant along the inflation trajectory
\begin{eqnarray}
\nonumber N&=&-\int_*^c\frac{pU}{M_p^2U_{,\varphi}}d\varphi-\int_*^c\frac{qV^{\nu+1}}{M_p^2V_{,\chi}}d\chi,\\
C&=&\int\frac{U^{\nu+1}}{M_p^2U_{,\varphi}}d\varphi-\int\frac{V}{M_p^2V_{,\chi}}d\chi.
\end{eqnarray}
Parallel to section \ref{sect-modelI}, we also calculated the
slow-roll parameters in this model,
\begin{eqnarray}\label{slrollII}
\nonumber &&\epsilon_{\varphi}=\frac{M_p^2}{2}\frac{U_{,\varphi}^2}{(p+qw^{\nu})^2U^2},~~~~\epsilon_{\chi}=\frac{M_p^2}{2}\frac{U^{\nu}V_{,\chi}^2}{(p+qw^{\nu})^2V^2},\\
\nonumber &&\epsilon_b=\frac{2\nu^2M_p^2U_{,\varphi}^2}{U^2},~~~~\epsilon=\frac{M_p^2}{2}\frac{U_{,\varphi}^2V^2+U^{\nu+2}V_{,\chi}^2}{(p+qw^{\nu})^2w^2},\\
\nonumber &&\eta_{\varphi\varphi}=\frac{M_p^2[1-p-q(\nu+1)w^{\nu}]U_{,\varphi}^2}{(p+qw^{\nu})^2U^2}+\frac{M_p^2U_{,\varphi\varphi}}{(p+qw^{\nu})U},\\
\nonumber &&\eta_{\varphi\chi}=\frac{M_p^2(1-q\nu w^{\nu})U^{\nu/2}U_{,\varphi}V_{,\chi}}{(p+qw^{\nu})^2w},\\
&&\eta_{\chi\chi}=\frac{M_p^2[1-p-q(\nu+1)w^{\nu}]U^{\nu}V_{,\chi}^2}{(p+qw^{\nu})^2V^2}+\frac{M_p^2U^{\nu}V_{,\chi\chi}}{(p+qw^{\nu})V}.
\end{eqnarray}

Subsequently, after obtaining the equations
\begin{eqnarray}
\nonumber &&\frac{d\varphi_c}{dC}=\left.\frac{M_p^2UU_{,\varphi}V_{,\chi}^2}{U_{,\varphi}^2V^2+U^{\nu+2}V_{,\chi}^2}\right|_c,\\
\nonumber &&\frac{d\chi_c}{dC}=-\left.\frac{M_p^2U_{,\varphi}^2VV_{,\chi}}{U_{,\varphi}^2V^2+U^{\nu+2}V_{,\chi}^2}\right|_c,\\
&&\frac{\partial C}{\partial\varphi_*}=\frac{U^{*\nu+1}}{M_p^2U^*_{,\varphi}},~~~~\frac{\partial C}{\partial\chi_*}=-\frac{V^*}{M_p^2V^*_{,\chi}}
\end{eqnarray}
and
\begin{eqnarray}
&&Z=\frac{qU_{,\varphi}^2V^{\nu+2}-pU^2V_{,\chi}^2}{U_{,\varphi}^2V^2+U^{\nu+2}V_{,\chi}^2},
\end{eqnarray}
we find by a little computation
\begin{eqnarray}
\nonumber \frac{M_pU^*_{,\varphi}}{\sqrt{2}(p+qw^{*\nu})U^*}\frac{\partial Z^c}{\partial\varphi_*}&=&-\frac{M_pU^{*\nu}V^*_{,\chi}}{\sqrt{2}(p+qw^{*\nu})V^*}\frac{\partial Z^c}{\partial\chi_*}\\
&=&\frac{\sqrt{2}(p+qw^{*\nu})\mathcal{A}}{M_pU^{*\nu}}.
\end{eqnarray}
Here notation $\mathcal{A}$ is different from the one in the
previous section,
\begin{eqnarray}
\nonumber \mathcal{A}&=&\frac{U^{*2\nu}(p+qw^{c\nu})^2}{U^{c2\nu}(p+qw^{*\nu})^2}\frac{\epsilon^c_{\varphi}\epsilon^c_{\chi}}{\epsilon^{c3}}[p\nu\epsilon^{c2}_{\chi}-q\nu w^{c\nu}\epsilon^{c2}_{\varphi}\\
&&-(4-2q\nu w^{c\nu})\epsilon^c_{\varphi}\epsilon^c_{\chi}+\epsilon^c_{\chi}\eta^c_{\varphi\varphi}+\epsilon^c_{\varphi}\eta^c_{\chi\chi}].
\end{eqnarray}

In terms of
\begin{equation}
u=\frac{p+Z^cU^{*\nu}}{p+qw^{*\nu}},~~~~v=\frac{qw^{*\nu}-Z^cU^{*\nu}}{p+qw^{*\nu}}
\end{equation}
and the relation $u+v=1$, once again straightforward calculation
gives
\begin{eqnarray}
\nonumber &&N_{,\varphi_*}=\frac{(p+qw^{*\nu})U^*u}{M_p^2U^*_{,\varphi}},~~~~N_{,\chi_*}=\frac{(p+qw^{*\nu})V^*v}{M_p^2U^{*\nu}V^*_{,\chi}},\\
\nonumber &&N_{,\varphi_*\varphi_*}=\frac{1}{M_p^2}\left[\left(1-\frac{\eta^*_{\varphi\varphi}}{2\epsilon^*_{\varphi}}\right)u-p\nu v+\frac{\mathcal{A}}{\epsilon^*_{\varphi}}\right],\\
\nonumber &&N_{,\varphi_*\chi_*}=-\frac{2(p+qw^{*\nu})^2w^*\mathcal{A}}{M_p^4U^{*\nu}U^*_{,\varphi}V^*_{,\chi}},\\
&&N_{,\chi_*\chi_*}=\frac{1}{M_p^2U^{*\nu}}\left[\left(1-\frac{\eta^*_{\chi\chi}}{2\epsilon^*_{\chi}}\right)v+q\nu w^{*\nu}u+\frac{\mathcal{A}}{\epsilon^*_{\chi}}\right].
\end{eqnarray}

Therefore, the non-linear parameter in this model is
\begin{eqnarray}\label{fNL4II}
\nonumber -\frac{6}{5}f_{\mathrm{NL}}^{(4)}&=&2\left(\frac{u^2}{\epsilon^*_{\varphi}}+\frac{v^2}{\epsilon^*_{\chi}}\right)^{-2}\biggl\{\frac{u^2}{\epsilon^*_{\varphi}}\left[\left(1-\frac{\eta^*_{\varphi\varphi}}{2\epsilon^*_{\varphi}}\right)u-p\nu v\right]\\
&&+\frac{v^2}{\epsilon^*_{\chi}}\left[\left(1-\frac{\eta^*_{\chi\chi}}{2\epsilon^*_{\chi}}\right)v+q\nu w^{*\nu}u\right]+\left(\frac{u}{\epsilon^*_{\varphi}}-\frac{v}{\epsilon^*_{\chi}}\right)^2\mathcal{A}\biggr\}.
\end{eqnarray}

Similar to the previous section, we can estimate
$f_{\mathrm{NL}}^{(4)}$ by assuming $u\sim v\sim\mathcal{O}(1)$ and
$\epsilon\sim\eta\ll1$. Under these assumptions, the only
possibility to generate a large non-linear parameter is in the limit
$\nu\gg1$. Unfortunately, careful analysis ruled out this
possibility. Because the assumption
$\epsilon_{\varphi}\sim\epsilon_b$ implies $p\nu+q\nu
w^{\nu}\sim\mathcal{O}(1)$, we find the non-linear parameter is not
enhanced by $\nu$ but is suppressed by the slow-roll parameters,
$f_{\mathrm{NL}}^{(4)}\sim\epsilon$. The same suppression applies if
$\nu$ lies in other regions. So we conclude that it is hopeless to
generate large non-Gaussianities in this model unless one goes
beyond the assumptions we made. A careful scan of parameter space
will be done in subsection \ref{subsect-nsPotIII} for a specific
model.

\section{Examples}\label{sect-examples}
In sections above, we have generalized the method of
\cite{Vernizzi:2006ve,Choi:2007su} and applied it to a larger class
of models. These models are summarized by equations \eqref{modelIa}
and \eqref{modelIIb}, whose non-linear parameters are given by
\eqref{fNL4I} and \eqref{fNL4II} generally. To check our general
formulae, we will reduce \eqref{fNL4I} and \eqref{fNL4II} to
previously known limit in subsections \ref{subsect-sumPot} and
\ref{subsect-prodPot}. The reduced expressions are consistent with
the results of \cite{Vernizzi:2006ve,Choi:2007su}. In subsections
\ref{subsect-nsPotI}, \ref{subsect-nsPotII} and
\ref{subsect-nsPotIII}, we will apply our formulae to non-separable
examples and scan the full parameter spaces.

We should stress that all results in this paper are reliable only in
the slow-roll region, that means at the least $\epsilon^*_i\ll1$,
$\epsilon^*_b\ll1$, $|\eta^*_{ij}|\ll1$ with $i,j=\varphi,\chi$. A
method free of slow-roll condition for some special models has been
explored in reference \cite{Byrnes:2009qy}.

\subsection{Additive Potential: $W=w$, $w=U(\varphi)+V(\chi)$, $b=0$}\label{subsect-sumPot}
This potential is obtained from \eqref{modelIa} by setting
$\alpha=1$. The condition $b=0$ is necessary to guarantee
\eqref{ansatzQ}. After taking $\alpha=1$, the result in section
\ref{sect-modelI} matches with that in \cite{Vernizzi:2006ve}
obviously.

\subsection{Multiplicative Potential: $W=w$, $w=U(\varphi)V(\chi)$}\label{subsect-prodPot}
Like equation \eqref{modelIIb2}, we leave $b$ as an arbitrary
function of $\varphi$, as long as the slow-roll parameters
\eqref{slroll} are small. This is a special limit of section
\ref{sect-modelII}.

Using relations
\begin{eqnarray}
\nonumber &&p=1,~~~~q=0,~~~~U^{\nu}=e^{-2b},\\
&&\nu=-\frac{1}{2}\sign(b_{,\varphi})\sign\left(\frac{U_{,\varphi}}{U}\right)\sqrt{\frac{\epsilon_b}{\epsilon_{\varphi}}},
\end{eqnarray}
we get the reduced form of non-linear parameter
\begin{eqnarray}
\nonumber -\frac{6}{5}f_{\mathrm{NL}}^{(4)}&=&2\left(\frac{u^2}{\epsilon^*_{\varphi}}+\frac{v^2}{\epsilon^*_{\chi}}\right)^{-2}\biggl[\frac{u^3}{\epsilon^*_{\varphi}}\left(1-\frac{\eta^*_{\varphi\varphi}}{2\epsilon^*_{\varphi}}\right)+\frac{v^3}{\epsilon^*_{\chi}}\left(1-\frac{\eta^*_{\chi\chi}}{2\epsilon^*_{\chi}}\right)\\
&&+\frac{u^2v}{2\epsilon^*_{\varphi}}\sign(b_{,\varphi})\sign\left(\frac{U_{,\varphi}}{U}\right)\sqrt{\frac{\epsilon^*_b}{\epsilon^*_{\varphi}}}+\left(\frac{u}{\epsilon^*_{\varphi}}-\frac{v}{\epsilon^*_{\chi}}\right)^2\mathcal{A}\biggr],
\end{eqnarray}
where we have made use of the fact that $u+v=1$ as well as the
following notations
\begin{equation}
u=1-\frac{\epsilon^c_{\chi}}{\epsilon^c}e^{2b_c-2b_*},~~~~v=\frac{\epsilon^c_{\chi}}{\epsilon^c}e^{2b_c-2b_*},
\end{equation}
\begin{eqnarray}
\nonumber \mathcal{A}&=&\frac{\epsilon^c_{\varphi}\epsilon^c_{\chi}}{\epsilon^{c3}}e^{4b_c-4b_*}\biggl[\epsilon^c_{\chi}\eta^c_{\varphi\varphi}+\epsilon^c_{\varphi}\eta^c_{\chi\chi}-4\epsilon^c_{\varphi}\epsilon^c_{\chi}\\
&&-\frac{1}{2}\sign(b_{,\varphi})\sign\left(\frac{U_{,\varphi}}{U}\right)\epsilon^{c2}_{\chi}\sqrt{\frac{\epsilon^*_b}{\epsilon^*_{\varphi}}}\biggr].
\end{eqnarray}

One may compare this formula with \cite{Choi:2007su}. Note that
their definitions of $u$, $v$ and $\mathcal{A}$ are slightly
different from ours by some factors. Taking these factors into
account, the result here is in accordance with \cite{Choi:2007su}.

\subsection{Non-separable Potential I: $W=\left(\alpha\varphi^2+\beta\chi^2\right)^{\nu}$, $b=0$}\label{subsect-nsPotI}
We spend an independent subsection on this model not because of its
non-Gaussianity, but because it has an elegant relation between the
$e$-folding number and the angle variable of fields. For this model,
the number of $e$-foldings from time $t$ during the inflation stage
to the end of inflation is
\begin{equation}
\ln\frac{a_e}{a(t)}=s(t)-s_e=\frac{\varphi^2+\chi^2}{4\nu M_p^2}-\frac{\varphi_e^2+\chi_e^2}{4\nu M_p^2}.
\end{equation}
Note that $\nu s$ can be regarded as sum of squares. Its time
derivative gives the Hubble parameter $ds/dt=-H$. So we can follow
the standard treatment to parameterize the scalars in polar
coordinates
\begin{equation}
\varphi=2M_p\sqrt{\nu s}\sin\theta,~~~~~\chi=2M_p\sqrt{\nu s}\cos\theta.
\end{equation}

Rewriting the equations of motion \eqref{eom} in terms of the polar
coordinates, we obtain a differential relation between $s$ and
$\theta$ for the present model,
\begin{equation}
\sin^2\theta+\frac{d\sin^2\theta}{d\ln(\nu s)}=\frac{R\sin^2\theta}{R\sin^2\theta+\cos^2\theta}
\end{equation}
with $R=\alpha/\beta$. It can be solved out to give
\begin{equation}\label{PSrelation}
N+s_e=s=s_0\frac{(\sin\theta)^{2/(R-1)}}{(\cos\theta)^{2R/(R-1)}}.
\end{equation}
At the end of inflation, if the scalars arrive at the bottom of
potential, one may simply set $s_e=0$.

Relation \eqref{PSrelation} is a trivial but useful generalization
of Polarski and Starobinsky's relation
\cite{Polarski:1992dq,Langlois:1999dw,Vernizzi:2006ve}. Recall that
Polarski and Starobinsky's relation has been widely used for the
inflation model with two massive scalar fields, which corresponds to
exponent $\nu=1$ in the model of this subsection. The simple
demonstration above generalized the relation to arbitrary $\nu$.

As an application, we evaluate \eqref{PSrelation} on the initial
flat hypersurface $t=t_*$ and then on the final comoving
hypersurface $t=t_c$, getting the ratio
\begin{equation}
\frac{s_c}{s_*}=\left(\frac{\sin\theta_c}{\sin\theta_*}\right)^{2/(R-1)}\left(\frac{\cos\theta_*}{\cos\theta_c}\right)^{2R/(R-1)},
\end{equation}
which reduces to
\begin{equation}\label{CstarcnsPotII}
\frac{\varphi_c^2}{\varphi_*^2}=\left(\frac{\chi_c^2}{\chi_*^2}\right)^R.
\end{equation}
This result can be also achieved from \eqref{Cstarc} directly.

\subsection{Non-separable Potential II: $W=\left(\alpha\varphi^2+\beta\chi^2\right)^2$, $b=0$}\label{subsect-nsPotII}
Our purpose in this and the next subsections is to examine
non-Gaussianities by parameter scanning. Two common assumptions will
be used: the $e$-folding number is fixed to be $N=60$ and the
inflation is supposed to conclude at the point
$\epsilon^c_{\varphi}+\epsilon^c_{\chi}=1$.

Using the latter assumption and the general formulae in section
\ref{sect-modelI}, we find all of the relevant quantities can be
expressed by $\epsilon_{\varphi}$, $\epsilon_{\chi}$ and $R$:
\begin{eqnarray}\label{phichinsPotII}
\nonumber \varphi^2=\frac{8M_p^2\epsilon_{\varphi}}{(\epsilon_{\varphi}+R\epsilon_{\chi})^2},&&\chi^2=\frac{8M_p^2R^2\epsilon_{\chi}}{(\epsilon_{\varphi}+R\epsilon_{\chi})^2},\\
\eta_{\varphi\varphi}=\frac{1}{2}(3\epsilon_{\varphi}+R\epsilon_{\chi}),&&\eta_{\chi\chi}=\frac{1}{2}\left(3\epsilon_{\chi}+\frac{\epsilon_{\varphi}}{R}\right),
\end{eqnarray}
\begin{eqnarray}
\nonumber u&=&\frac{\epsilon^*_{\varphi}}{\epsilon^*_{\varphi}+R\epsilon^*_{\chi}}+\frac{(R-1)\epsilon^c_{\varphi}\epsilon^c_{\chi}(\epsilon^*_{\varphi}+R\epsilon^*_{\chi})}{(\epsilon^c_{\varphi}+R\epsilon^c_{\chi})^2},\\
\nonumber v&=&\frac{R\epsilon^*_{\chi}}{\epsilon^*_{\varphi}+R\epsilon^*_{\chi}}-\frac{(R-1)\epsilon^c_{\varphi}\epsilon^c_{\chi}(\epsilon^*_{\varphi}+R\epsilon^*_{\chi})}{(\epsilon^c_{\varphi}+R\epsilon^c_{\chi})^2},\\
\mathcal{A}&=&-\frac{\epsilon^c_{\varphi}\epsilon^c_{\chi}(\epsilon^*_{\varphi}+R\epsilon^*_{\chi})^2}{4(\epsilon^c_{\varphi}+R\epsilon^c_{\chi})^2}\left[1-\frac{(\epsilon^c_{\varphi}+R\epsilon^c_{\chi})^2}{R}\right],
\end{eqnarray}
\begin{eqnarray}
\nonumber -\frac{6}{5}f_{\mathrm{NL}}^{(4)}&=&(\epsilon^*_{\chi}u^2+\epsilon^*_{\varphi}v^2)^{-2}\biggl[\frac{1}{2}\epsilon^{*2}_{\chi}u^3(\epsilon^*_{\varphi}-R\epsilon^*_{\chi})+\frac{1}{2}\epsilon^{*2}_{\varphi}v^3\left(\epsilon^*_{\chi}-\frac{\epsilon^*_{\varphi}}{R}\right)\\
&&+\epsilon^*_{\varphi}\epsilon^*_{\chi}uv(\epsilon^*_{\chi}u+\epsilon^*_{\varphi}v)+4\mathcal{A}(\epsilon^*_{\chi}u+\epsilon^*_{\varphi}v)^2\biggr],
\end{eqnarray}
\begin{eqnarray}\label{Nstar}
\nonumber N&=&\frac{\varphi_*^2+\chi_*^2}{8M_p^2}-\frac{\varphi_c^2+\chi_c^2}{8M_p^2}\\
&=&\frac{\epsilon^*_{\varphi}+R^2\epsilon^*_{\chi}}{(\epsilon^*_{\varphi}+R\epsilon^*_{\chi})^2}-\frac{\epsilon^c_{\varphi}+R^2\epsilon^c_{\chi}}{(\epsilon^c_{\varphi}+R\epsilon^c_{\chi})^2}.
\end{eqnarray}
Here we defined $R=\alpha/\beta$ like the previous subsection. If
$R=1$, it can be proved that
$-6f_{\mathrm{NL}}^{(4)}/5=(\epsilon^*_{\varphi}+\epsilon^*_{\chi})/2=1/(2N+2)$.
Without loss of generality, we will consider the parameter region
$0<R\leq1$. As has been mentioned, from \eqref{Cstarc} or
\eqref{PSrelation}, one can get relation \eqref{CstarcnsPotII}. This
relation is equivalent to
\begin{equation}\label{CstarcsrnsPotII}
\frac{\epsilon^c_{\varphi}}{\epsilon^*_{\varphi}}=\left(\frac{\epsilon^c_{\chi}}{\epsilon^*_{\chi}}\right)^R\left(\frac{\epsilon^c_{\varphi}+R\epsilon^c_{\chi}}{\epsilon^*_{\varphi}+R\epsilon^*_{\chi}}\right)^{2(1-R)}.
\end{equation}
If $R=1$, it gives
$\epsilon^c_{\varphi}/\epsilon^*_{\varphi}=\epsilon^c_{\chi}/\epsilon^*_{\chi}=1/(\epsilon^*_{\varphi}+\epsilon^*_{\chi})=N+1$
and thus $\eta^*_{\chi\chi}=(2\epsilon^c_{\chi}+1)/(2N+2)$.

In the above expressions, there are five parameters:
$\epsilon^*_{\varphi}$, $\epsilon^*_{\chi}$, $\epsilon^c_{\varphi}$,
$\epsilon^c_{\chi}$ and $R$. The number can be reduced by the
assumptions we made at the beginning of this section.\footnote{We
are very grateful to Christian T. Byrnes for pointing out an error
on this issue in an earlier version.} Firstly,
$\epsilon^c_{\varphi}$ and $\epsilon^c_{\chi}$ can be traded to each
other with the relation $\epsilon^c_{\varphi}+\epsilon^c_{\chi}=1$.
Secondly, since we have assumed $N=60$, equations \eqref{Nstar} and
\eqref{CstarcsrnsPotII} can be used to eliminate two degrees of
freedom further. Now we see only two parameters are independent, and
we choose them to be $\epsilon^c_{\chi}$ and $R$ in the analysis
below. The number counting in this way agrees with the fact that
\eqref{eom} is a first order system under the slow-roll
approximation.

As a useful trick, we introduce a dimensionless notation
$x=\chi_*^2/\chi_c^2$, then equations \eqref{CstarcnsPotII} and
\eqref{Nstar} can be reformed as $\varphi_*^2/\varphi_c^2=x^R$ and
\begin{equation}
\frac{\varphi_c^2x^R+\chi_c^2x}{8M_p^2}=N+\frac{\varphi_c^2+\chi_c^2}{8M_p^2}.
\end{equation}
Usually the second equation has no analytical expression for the
root $x$, but one may still find the root numerically. In the region
$x>0$, both $x$ and $x^R$ increase monotonically from zero to
infinity, so this equation with respect to $x$ has exactly one
positive real root if the right hand side is finite. In terms of
$\epsilon^c_{\varphi}$, $\epsilon^c_{\chi}$ and $R$, this equation
is of the form
\begin{equation}\label{x}
\frac{\epsilon^c_{\varphi}x^R+R^2\epsilon^c_{\chi}x}{(\epsilon^c_{\varphi}+R\epsilon^c_{\chi})^2}=N+\frac{\epsilon^c_{\varphi}+R^2\epsilon^c_{\chi}}{(\epsilon^c_{\varphi}+R\epsilon^c_{\chi})^2}.
\end{equation}

\begin{figure}
\begin{center}
\includegraphics[width=0.45\textwidth]{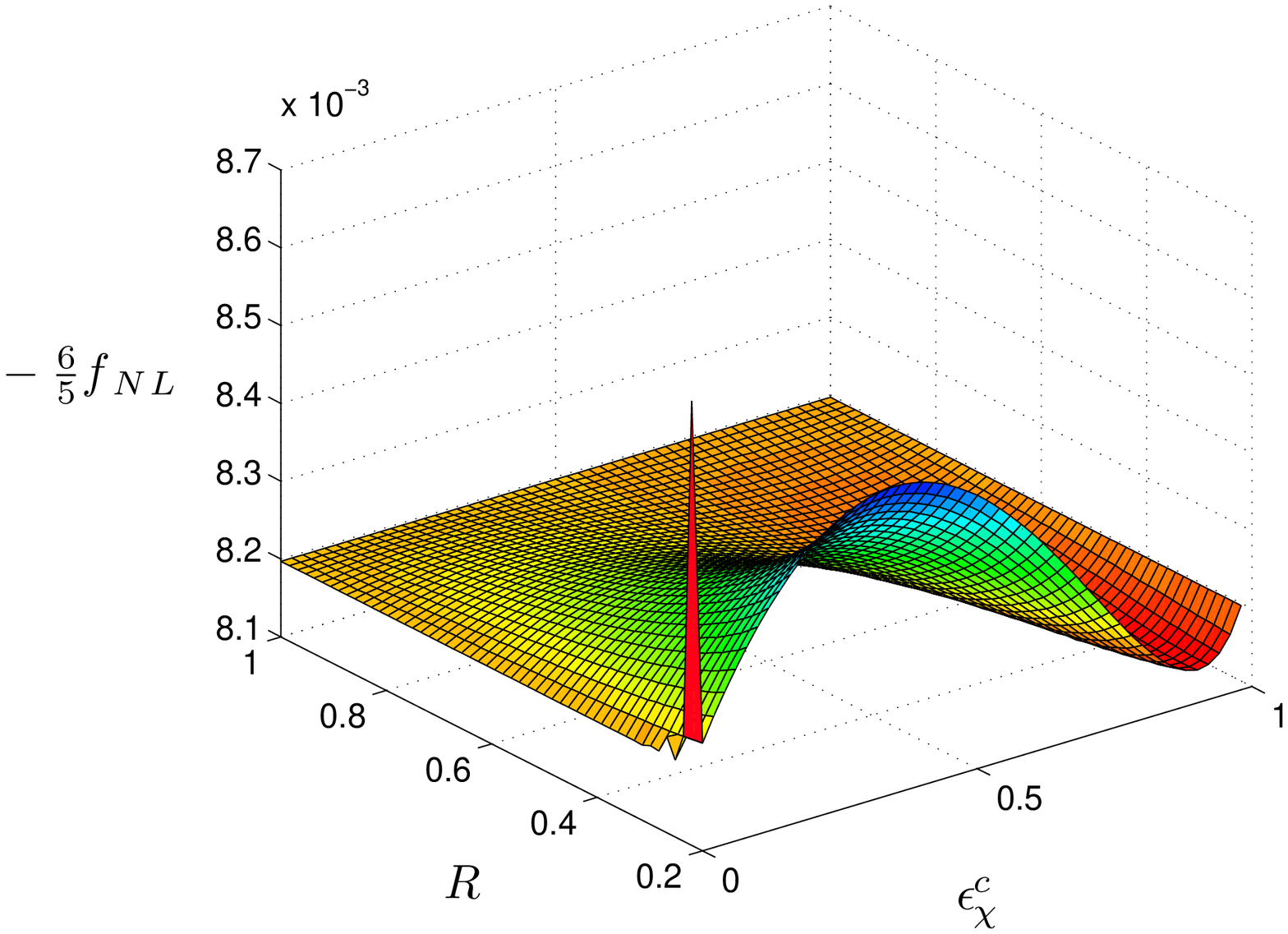}\\
\includegraphics[width=0.45\textwidth]{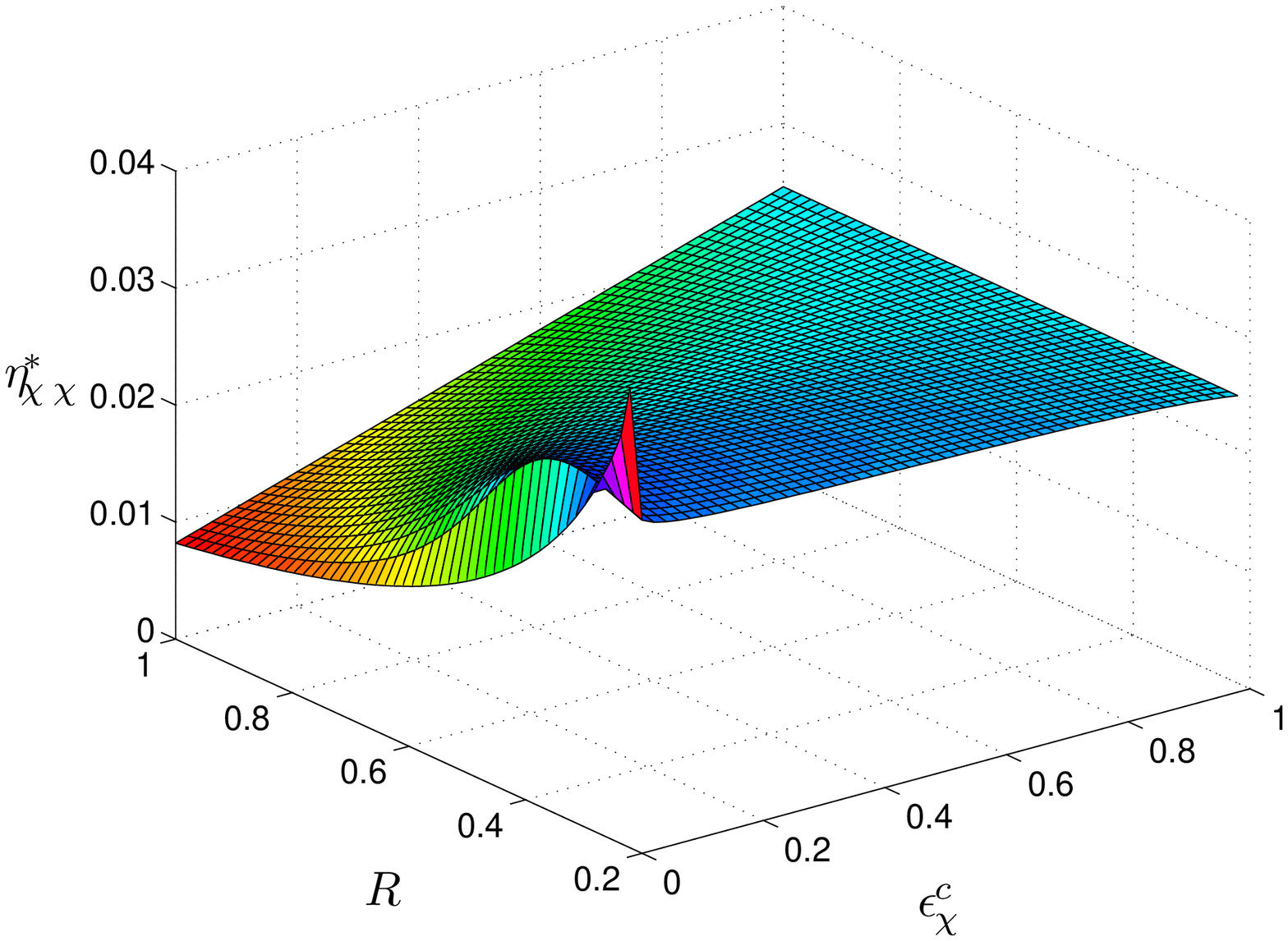}\\
\end{center}
\caption{\textbf{(color online). The non-linear parameter
\eqref{fNL4nsPotII} and slow-roll parameter $\eta^*_{\chi\chi}$
\eqref{phichinsPotII} as functions of $\epsilon^c_{\chi}$ and $R$,
under the assumptions $N=60$ and
$\epsilon^c_{\varphi}+\epsilon^c_{\chi}=1$. $R$ is defined as
$R=\alpha/\beta$, the ratio of two parameters in the potential of
this model.}}\label{fig-nsPotII}
\end{figure}

Fixing $N=60$, the recipe of our numerical simulation is as follows:
\begin{enumerate}
\item Given the values of $\epsilon^c_{\chi}$ and $R$ in parameter space
$0\leq\epsilon^c_{\chi}\leq1$, $0<R\leq1$, numerically find the root
$x$ of equation \eqref{x}, where
$\epsilon^c_{\varphi}=1-\epsilon^c_{\chi}$.
\item Compute $\epsilon^*_{\varphi}$, $\epsilon^*_{\chi}$,
$\eta^*_{\varphi\varphi}$ and $\eta^*_{\chi\chi}$ according to
\begin{equation}
\epsilon^*_{\varphi}=\frac{\epsilon^c_{\varphi}x^R(\epsilon^c_{\varphi}+R\epsilon^c_{\chi})^2}{(\epsilon^c_{\varphi}x^R+R\epsilon^c_{\chi}x)^2},~~~~\epsilon^*_{\chi}=\frac{\epsilon^c_{\chi}x(\epsilon^c_{\varphi}+R\epsilon^c_{\chi})^2}{(\epsilon^c_{\varphi}x^R+R\epsilon^c_{\chi}x)^2}
\end{equation}
and equations \eqref{phichinsPotII}.
\item Evaluate $-6f_{\mathrm{NL}}^{(4)}/5$ with the formula
\begin{eqnarray}\label{fNL4nsPotII}
\nonumber -\frac{6}{5}f_{\mathrm{NL}}^{(4)}&=&(\epsilon^c_{\varphi}+R\epsilon^c_{\chi})^2\left\{\frac{\epsilon^c_{\varphi}}{x^R}[x^R+(R-1)\epsilon^c_{\chi}]^2+\frac{\epsilon^c_{\chi}}{x}[Rx-(R-1)\epsilon^c_{\varphi}]^2\right\}^{-2}\\
\nonumber &&\times\Biggl\{\frac{\epsilon^c_{\varphi}}{2x^R}[x^R+(R-1)\epsilon^c_{\chi}]^2\left[1-\frac{(R-1)\epsilon^c_{\chi}}{x^R}\right]\\
\nonumber &&+\frac{\epsilon^c_{\chi}}{2x}[Rx-(R-1)\epsilon^c_{\varphi}]^2\left[1+\frac{(R-1)\epsilon^c_{\varphi}}{Rx}\right]\\
&&-\epsilon^c_{\varphi}\epsilon^c_{\chi}\left[1-\frac{(\epsilon^c_{\varphi}+R\epsilon^c_{\chi})^2}{R}\right]\left[\frac{x^R+(R-1)\epsilon^c_{\chi}}{x^R}+\frac{Rx-(R-1)\epsilon^c_{\varphi}}{x}\right]^2\Biggr\}.
\end{eqnarray}
\item Repeat the above steps to scan the entire parameter space of
$\epsilon^c_{\chi}$ and $R$. Due to the violation of slow-roll
condition, the vicinity of $R=0$ should be skipped to avoid numerical
singularities (see spikes in figure \ref{fig-nsPotII}).
\end{enumerate}

In a practical simulation, we scan the region
$0\leq\epsilon^c_{\chi}\leq1$, $0.001\leq R\leq1$ on a uniform grid
with $101^2$ points. Some simulation results are illustrated in
figure \ref{fig-nsPotII}. When drawing the figure, we have imposed
the slow-roll condition $\epsilon^*_{\varphi}<0.05$,
$\epsilon^*_{\chi}<0.05$, $\eta^*_{\varphi\varphi}<0.05$,
$\eta^*_{\chi\chi}<0.05$. In the limit $R=1$, they are in agreement
with the analytical results $-6f_{\mathrm{NL}}^{(4)}/5=1/(2N+2)$,
$\eta^*_{\chi\chi}=(2\epsilon^c_{\chi}+1)/(2N+2)$. One may also
check the results in other limits analytically, such as
$\epsilon^c_{\chi}\rightarrow0$ or $\epsilon^c_{\chi}\rightarrow1$.
Theoretically, $R\rightarrow0$ should correspond to an inflation
model driven by one field $\varphi$. But our method does not apply
to that limit, because it would violate the slow-roll condition for
$\chi$.

From figure \ref{fig-nsPotII},we can see the non-linear parameter
$f_{\mathrm{NL}}^{(4)}$ is suppressed by slow-roll parameters.
Especially, in the neighborhood of $R=0$, the spikes of
$f_{\mathrm{NL}}^{(4)}$ are located at the same positions as the spikes
of $\eta^*_{\chi\chi}$. Such a coincidence continues to exist even
if one relaxes the slow-roll condition. But there is no spike in
similar graphs for $\epsilon^*_{\varphi}$, $\epsilon^*_{\chi}$ and
$\eta^*_{\varphi\varphi}$. Actually, these spikes are mainly
attributed to the enhancement of $f_{\mathrm{NL}}^{(4)}$ and
$\eta^*_{\chi\chi}$ by $1/R$ in the small $R$ limit. After the
parameter scanning and the numerical simulation, our lesson is that
this model cannot generate a large non-Gaussianity unless the
slow-roll condition breaks down.

\subsection{Non-separable Potential III: $W=\lambda e^{-\beta\varphi^2\chi^2}$, $e^{2b}=\alpha\varphi^2$}\label{subsect-nsPotIII}
This is a special model of \eqref{modelIIb} with $p=0$, $\nu=-1$,
$q=-1$. As in the previous subsection, we assume $N=60$ and
$\epsilon^c_{\varphi}+\epsilon^c_{\chi}=1$. Then from section
\ref{sect-modelII} we get the relations
\begin{eqnarray}\label{slroll-nsPotIII}
\nonumber &&\frac{\epsilon_{\varphi}}{\epsilon_{\chi}}=\alpha\chi^2,~~~~\frac{\epsilon_{\chi}^2}{\epsilon_{\varphi}}=\frac{2M_p^2\beta^2}{\alpha^2}\varphi^2,\\
\nonumber &&\frac{4M_p^2\beta}{\alpha}=\frac{1}{N}\ln\left(\frac{\epsilon^c_{\varphi}\epsilon^*_{\chi}}{\epsilon^c_{\chi}\epsilon^*_{\varphi}}\right)=\epsilon_{\chi}\sqrt{\frac{\epsilon_b}{\epsilon_{\varphi}}},\\
&&\frac{1}{2}\sqrt{\frac{\epsilon_b}{\epsilon_{\varphi}}}=2-\frac{\eta_{\chi\chi}}{\epsilon_{\chi}}=2-\frac{\eta_{\varphi\varphi}}{\epsilon_{\varphi}}=1-\frac{\eta_{\varphi\chi}}{2\sqrt{\epsilon_{\varphi}\epsilon_{\chi}}},
\end{eqnarray}
\begin{eqnarray}
\nonumber u&=&1-v=\frac{\epsilon^c_{\chi}\epsilon^*_{\varphi}}{\epsilon^*_{\chi}},\\
\mathcal{A}&=&-\frac{\epsilon^{*2}_{\varphi}\epsilon^c_{\varphi}\epsilon^{c2}_{\chi}}{2\epsilon^{*2}_{\chi}}\frac{4M_p^2\beta}{\alpha},
\end{eqnarray}
\begin{equation}
-\frac{6}{5}f_{\mathrm{NL}}^{(4)}=(\epsilon^*_{\chi}u^2+\epsilon^*_{\varphi}v^2)^{-2}\biggl[\frac{\epsilon^*_{\varphi}}{2}\frac{4M_p^2\beta}{\alpha}(\epsilon^*_{\chi}u^3+\epsilon^*_{\varphi}v^3+2\epsilon^*_{\varphi}uv^2)+2\mathcal{A}(\epsilon^*_{\chi}u-\epsilon^*_{\varphi}v)^2\biggr].
\end{equation}

For the present model, equation \eqref{Cstarc} gives
\begin{equation}
\ln\left(\frac{\varphi_c^2}{\varphi_*^2}\right)=\alpha(\chi_c^2-\chi_*^2),
\end{equation}
that is
\begin{equation}\label{CstarcsrnsPotIII}
\ln\left(\frac{\epsilon^{c2}_{\chi}\epsilon^*_{\varphi}}{\epsilon^c_{\varphi}\epsilon^{*2}_{\chi}}\right)=\frac{\epsilon^c_{\varphi}}{\epsilon^c_{\chi}}-\frac{\epsilon^*_{\varphi}}{\epsilon^*_{\chi}}.
\end{equation}

If we introduce the notations
$R=(\epsilon^c_{\chi}\epsilon^*_{\varphi})/(\epsilon^c_{\varphi}\epsilon^*_{\chi})$,
then combining it with equation \eqref{CstarcsrnsPotIII} and the
condition $\epsilon^c_{\varphi}+\epsilon^c_{\chi}=1$, we can express
$\epsilon^*_{\varphi}$, $\epsilon^*_{\chi}$ and
$f_{\mathrm{NL}}^{(4)}$ in terms of $\epsilon^c_{\varphi}$,
$\epsilon^c_{\chi}$ and $R$,
\begin{eqnarray}
\nonumber \epsilon^*_{\varphi}&=&R^2\epsilon^c_{\varphi}\exp\left[\frac{(R-1)\epsilon^c_{\varphi}}{\epsilon^c_{\chi}}\right],\\
\epsilon^*_{\chi}&=&R\epsilon^c_{\chi}\exp\left[\frac{(R-1)\epsilon^c_{\varphi}}{\epsilon^c_{\chi}}\right],
\end{eqnarray}
\begin{equation}\label{fNL4nsPotIII}
-\frac{6}{5}f_{\mathrm{NL}}^{(4)}=\frac{1}{N}\ln\left(\frac{1}{R}\right)\frac{1-R\epsilon^c_{\varphi}+R^2(R-1)\epsilon^{c3}_{\varphi}-2R^2(R-1)^2\epsilon^{c5}_{\varphi}}{2[1-R\epsilon^c_{\varphi}+R(R-1)\epsilon^{c2}_{\varphi}]^2}.
\end{equation}
On the basis of equation \eqref{slroll-nsPotIII}, we deduce that
$\ln(1/R)/N$ should be positive and suppressed by slow-roll
parameters. In particular,
\begin{eqnarray}
\label{etaphisnsPotIII}\eta^*_{\varphi\varphi}&=&2\epsilon^*_{\varphi}-\frac{R\epsilon^c_{\varphi}}{\epsilon^c_{\chi}}\frac{1}{N}\ln\left(\frac{1}{R}\right),\\
\label{etachisnsPotIII}\eta^*_{\chi\chi}&=&2\epsilon^*_{\chi}-\frac{1}{N}\ln\left(\frac{1}{R}\right).
\end{eqnarray}
Thus we focus on the region $0<R<1$.

As indicated by the above analysis, if we are interested only in the
non-linear parameter and slow-roll parameters, this model has two
free parameters after using our assumptions and equations of motion.
They will be chosen as $\epsilon^c_{\chi}$ and $R$ in our
simulation, just like in the previous subsection. But we should warn
that, compared with the previous subsection, the notation $R$ has a
distinct meaning in the current subsection.

\begin{figure}
\begin{center}
\includegraphics[width=0.45\textwidth]{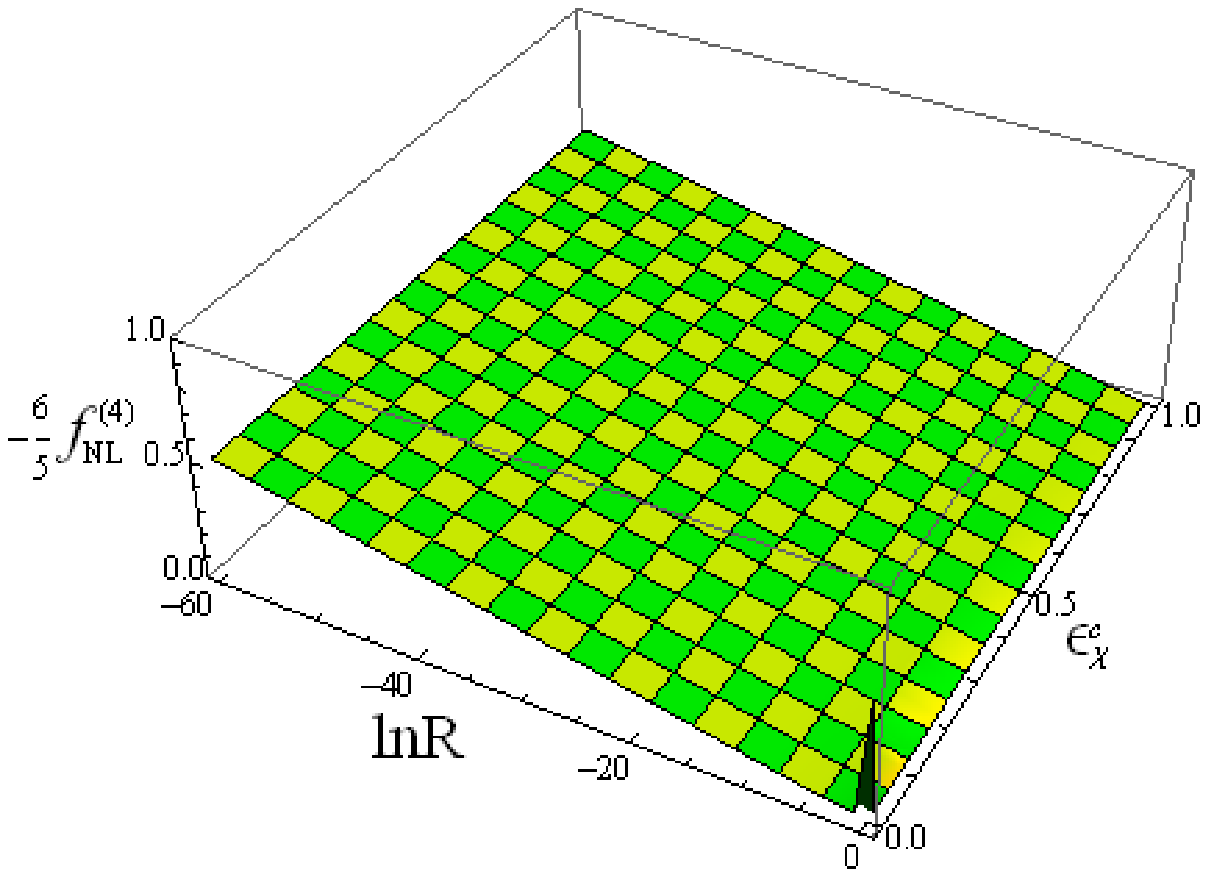}\\
\includegraphics[width=0.45\textwidth]{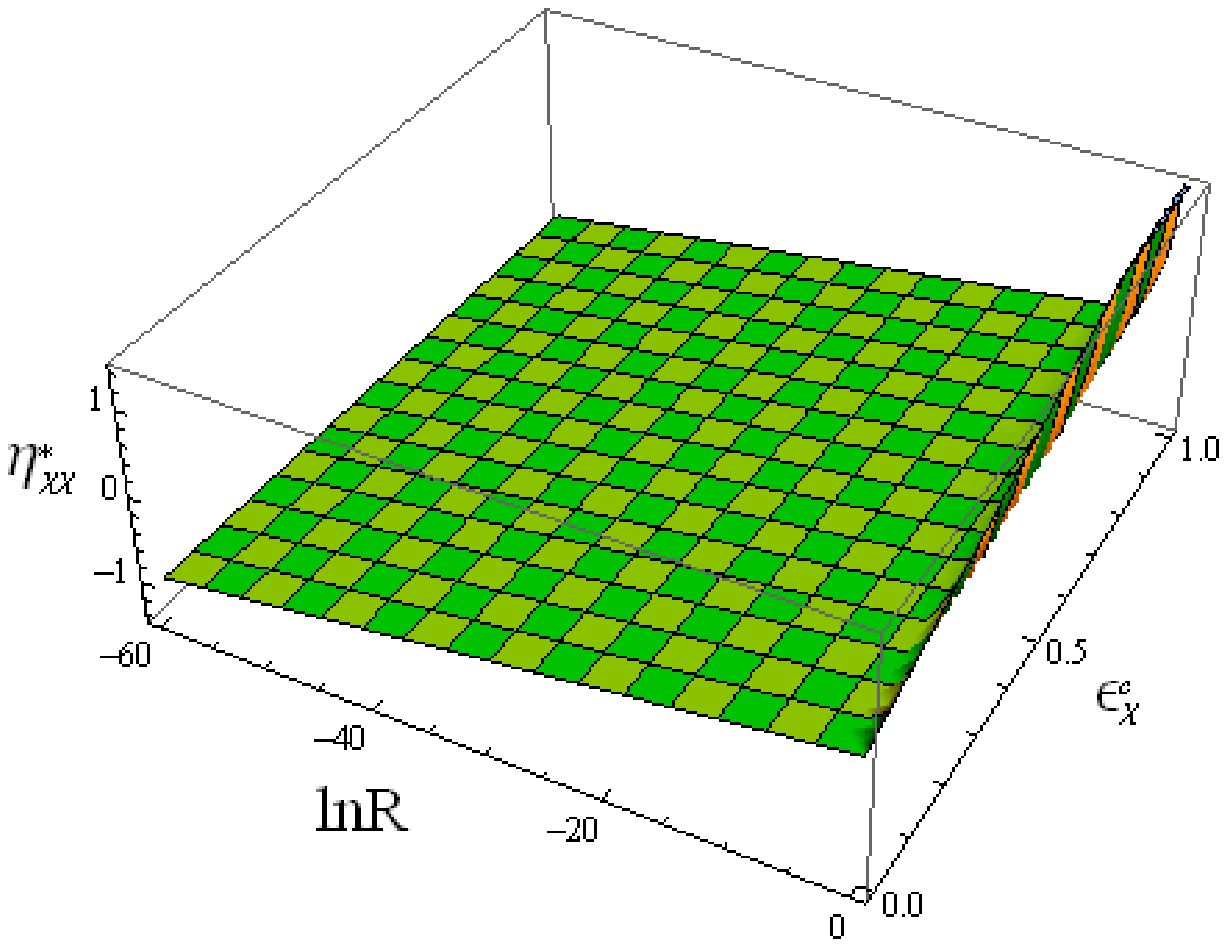}\\
\end{center}
\caption{\textbf{(color online). The non-linear parameter
\eqref{fNL4nsPotIII} and slow-roll parameter $\eta^*_{\chi\chi}$
\eqref{etachisnsPotIII} as functions of $\epsilon^c_{\chi}$ and $R$,
under the assumptions $N=60$ and
$\epsilon^c_{\varphi}+\epsilon^c_{\chi}=1$. $R$ is defined as
$R=(\epsilon^c_{\chi}\epsilon^*_{\varphi})/(\epsilon^c_{\varphi}\epsilon^*_{\chi})$,
and it is plotted in logarithmic scale.}}\label{fig-nsPotIIlog}
\end{figure}
\begin{figure}
\begin{center}
\includegraphics[width=0.45\textwidth]{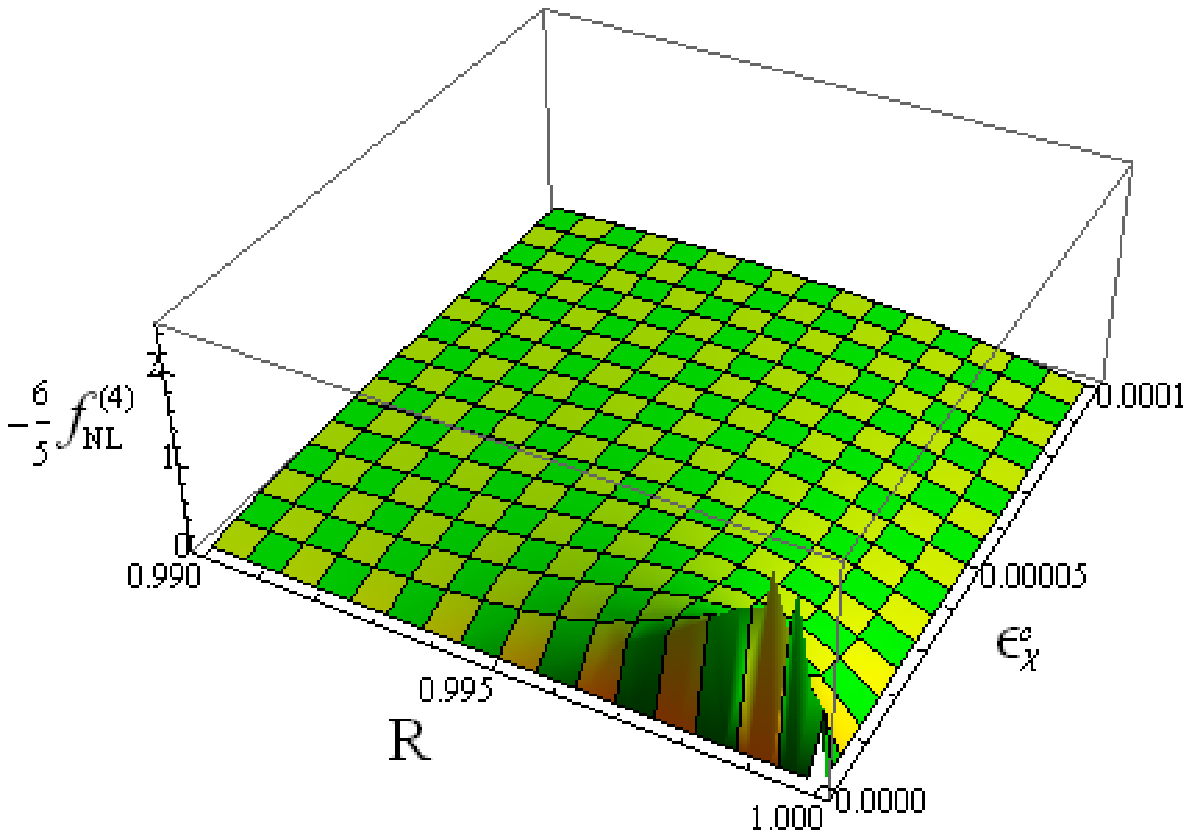}\\
\includegraphics[width=0.45\textwidth]{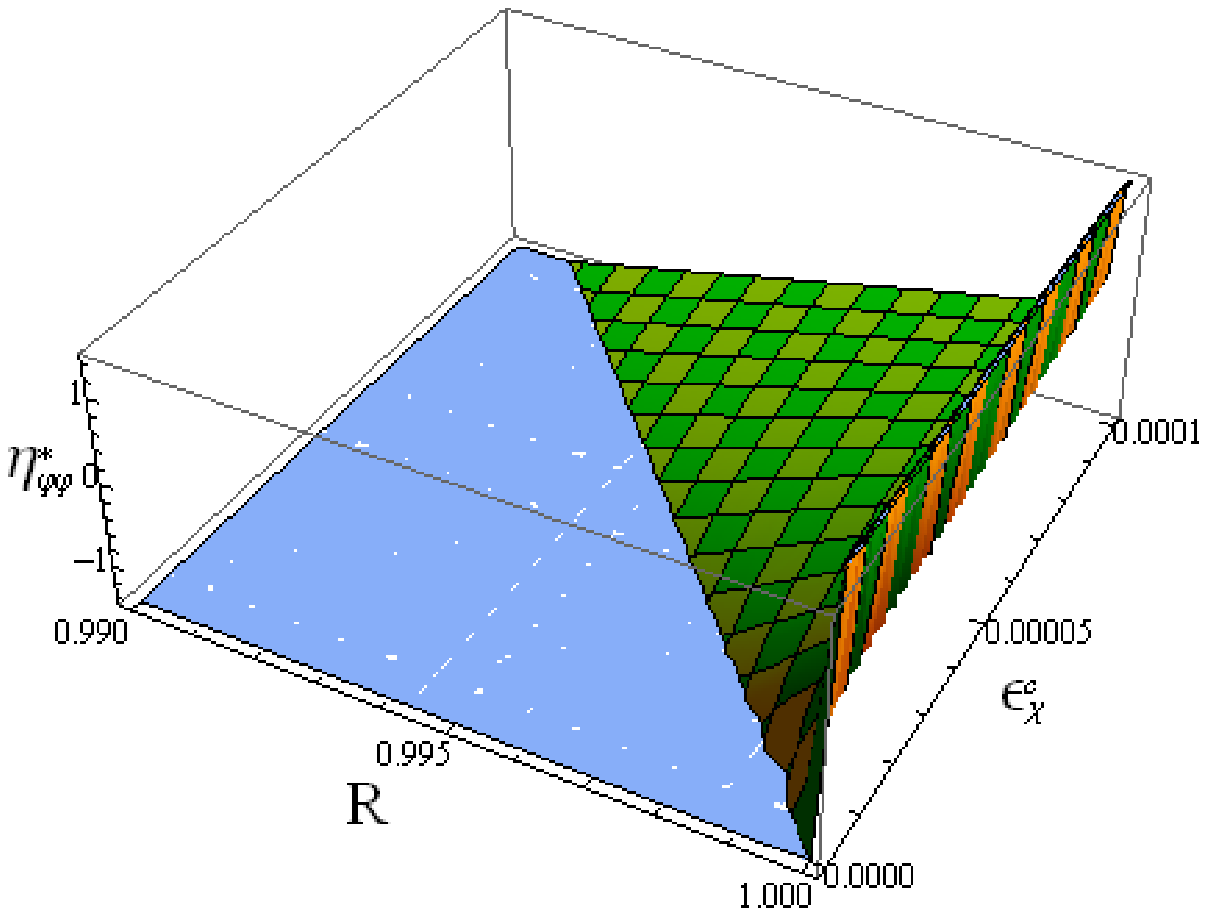}\\
\includegraphics[width=0.45\textwidth]{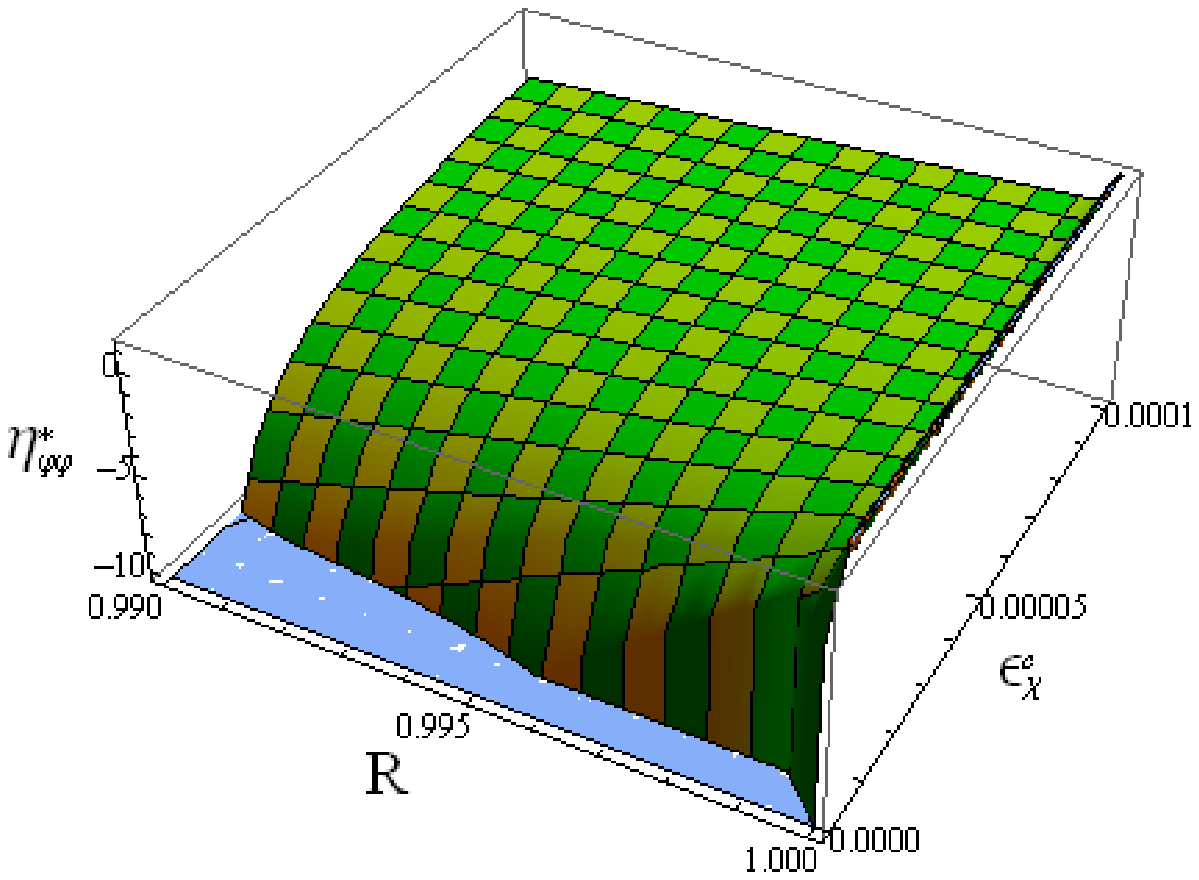}\\
\end{center}
\caption{\textbf{(color online). The non-linear parameter
\eqref{fNL4nsPotIII} and slow-roll parameter
$\eta^*_{\varphi\varphi}$ \eqref{etaphisnsPotIII} as functions of
$\epsilon^c_{\chi}$ and $R$ near the corner
$\epsilon^c_{\chi}\rightarrow0$, $R\rightarrow1$, under the
assumptions $N=60$ and $\epsilon^c_{\varphi}+\epsilon^c_{\chi}=1$.
$R$ is defined as
$R=(\epsilon^c_{\chi}\epsilon^*_{\varphi})/(\epsilon^c_{\varphi}\epsilon^*_{\chi})$,
and it is plotted in linear scale. In the middle and the lower
graphs, the regions with $\eta^*_{\varphi\varphi}<-1.5$ and
$\eta^*_{\varphi\varphi}<-10.5$ respectively are cut
off.}}\label{fig-nsPotIIlin}
\end{figure}

The parameter scanning is illustrated by figures
\ref{fig-nsPotIIlog} and \ref{fig-nsPotIIlin}. In figure
\ref{fig-nsPotIIlog}, parameter $R$ decreases exponentially from 1
to $e^{-60}$. In this process, the non-linear parameter grows
roughly proportional to  $\ln(1/R)$ while the slow-roll condition
$|\eta^*_{\chi\chi}|\ll1$ is violated gradually. This phenomenon
agrees with equations \eqref{etachisnsPotIII} and
\eqref{fNL4nsPotIII}, both of whose amplitude are enhanced by the
factor $\ln(1/R)/N$ when $R$ is small. In figure
\ref{fig-nsPotIIlog},we find a sharp spike for the non-linear
parameter in the corner $\epsilon^c_{\chi}\rightarrow0$,
$R\rightarrow1$. Figure \ref{fig-nsPotIIlin} is drawn to zoom in
this corner, with $R$ scaled linearly. As shown by this figure, the
spike dwells in a position violating the slow-roll condition
$|\eta^*_{\varphi\varphi}|\ll1$. Therefore, the non-linear parameter
in this model must be small once the slow-roll condition
$\epsilon^*_i\ll1$, $\epsilon^*_b\ll1$, $|\eta^*_{ij}|\ll1$
($i,j=\varphi,\chi$) is imposed.

\section{Summary}\label{sect-concl}
In this paper, we investigated a class of two-field slow-roll
inflation models whose non-linear parameter is analytically
calculable.

In our convention of notations, we collected some well-known but
necessary knowledge in section \ref{sect-preparation}. Slightly
generalizing the method of \cite{Vernizzi:2006ve,Choi:2007su}, we
showed in section \ref{sect-hunt} how their method could be utilized
in a larger class of models satisfying two ansatzes, namely
\eqref{ansatzf} and \eqref{ansatzQ}. In subsections
\ref{subsect-caseI} and \ref{subsect-caseII} we proposed models
meeting these ansatzes. We put our models in the form of $W(w)$ with
$w=U(\varphi)+V(\chi)$ in subsection \ref{subsect-caseI} and with
$w=U(\varphi)V(\chi)$ in subsection \ref{subsect-caseII}. At first
glance, these are two different classes of models. But in fact they
are two dual forms of the same class of models, just as proved in
subsection \ref{subsect-IeqII}. In a succinct form, our models can
be summarized by equations \eqref{modelIa} and \eqref{modelIIb},
whose non-linear parameters were worked out in sections
\ref{sect-modelI} and \ref{sect-modelII} respectively, see equations
\eqref{fNL4I} and \eqref{fNL4II}. Under simplistic assumptions, we
found no large non-Gaussianity in these models.

As a double check, we reduced the expression \eqref{fNL4I} for
non-linear parameter to the additive potential in subsection
\ref{subsect-sumPot}, and \eqref{fNL4II} to multiplicative potential
in subsection \ref{subsect-prodPot}. The resulting non-linear
parameters match with \cite{Vernizzi:2006ve,Choi:2007su}, confirming
our calculations. In subsection \ref{subsect-nsPotI}, for a special
class of models, we generalized Polarski and Starobinsky's relation
\eqref{PSrelation}. For more specific models, we scanned the
parameter space to evaluate the non-linear parameter, as shown by
figures in subsections \ref{subsect-nsPotII} and
\ref{subsect-nsPotIII}. In the scanning, we assumed the $e$-folding
number $N=60$ and the inflation terminates at $-\dot{H}/H^2=1$. For
the models we studied in subsections \ref{subsect-nsPotII} and
\ref{subsect-nsPotIII}, the non-linear parameter
$-6f_{\mathrm{NL}}^{(4)}/5$ always takes a small positive value
under the slow-roll approximation.

\acknowledgments{The author would like to thank Christian T. Byrnes for
private communications and helpful comments .}

\end{document}